\documentclass[fleqn,usenatbib]{mnras}

\usepackage{newtxtext,newtxmath}

\usepackage[T1]{fontenc}

\DeclareRobustCommand{\VAN}[3]{#2}
\let\VANthebibliography\thebibliography
\def\thebibliography{\DeclareRobustCommand{\VAN}[3]{##3}\VANthebibliography}

\usepackage{graphicx}	
\usepackage{amsmath}	
\usepackage{mathtools}
\usepackage{siunitx}    
\usepackage{multirow}

\DeclarePairedDelimiter\abs{\lvert}{\rvert}%

\DeclareSIUnit \h {h}
\DeclareSIUnit \parsec {pc}

\title[Wavelet analysis on high-redshift quasars]{Improving IGM temperature constraints using wavelet analysis on high-redshift quasars}

\author[Molly Wolfson et al.]{Molly Wolfson$^{1}$\thanks{E-mail: mawolfson@ucsb.edu},
Joseph F. Hennawi$^{1,2}$,
Frederick B. Davies$^{1,3,4}$,
Jose O{\~{n}}orbe$^{5}$,
Hector Hiss$^{4}$, 
\newauthor
and Zarija Luki\'{c}$^{3}$
\\
$^{1}$Department of Physics, University of California, Santa Barbara, CA 93106, USA\\
$^{2}$Leiden Observatory, Leiden University, Niels Bohrweg 2, 2333 CA Leiden, Netherlands\\
$^{3}$Lawrence Berkeley National Laboratory, 1 Cyclotron Rd, Berkeley, CA 94720, USA\\
$^{4}$Max-Planck-Institut f\"{u}r Astronomie, K\"{o}nigstuhl 17, 69117 Heidelberg, Germany\\
$^{5}$Facultad de F\'{i}sicas, Universidad de Sevilla, Avda. Reina Mercedes s/n, Campus de Reina Mercedes, 41012 Sevilla, Spain
}

\date{Accepted XXX. Received YYY; in original form ZZZ}

\pubyear{2020}

\begin{document}
\label{firstpage}
\pagerange{\pageref{firstpage}--\pageref{lastpage}}
\maketitle

\begin{abstract}
The thermal state of the intergalactic medium (IGM) contains vital information about the epoch of reionization, one of the most transformative yet poorly understood periods in the young universe.
This thermal state is encoded in the small-scale structure of Lyman-$\alpha$ (Ly$\alpha$) absorption in quasar spectra.
The 1D flux power spectrum measures the average small-scale structure along quasar sightlines. 
At high redshifts, where the opacity is large, averaging mixes high signal-to-noise ratio transmission spikes with noisy absorption troughs.  
Wavelet amplitudes are an alternate statistic that maintains spatial information while quantifying fluctuations at the same spatial frequencies as the power spectrum, giving them the potential to more sensitively measure the small-scale structure. 
Previous Ly$\alpha$ forest studies using wavelet amplitude probability density functions (PDFs) used limited spatial frequencies and neglected strong correlations between PDF bins and across wavelets scales, resulting in sub-optimal and unreliable parameter inference.
Here we present a novel method for performing statistical inference using wavelet amplitude PDFs that spans the full range of spatial frequencies probed by the power spectrum and that fully accounts for these correlations. 
We applied this procedure to realistic mock data drawn from a simple thermal model parameterized by the temperature at mean density, $T_0$, and find that wavelets deliver 1$\sigma$ constraints on $T_0$ that are on average 7\% more sensitive at $z=5$ (12\% at $z=6$) than those from the power spectrum. 
We consider the possibility of combing wavelet PDFs with the power, but find that this does not lead to improved sensitivity. 
\end{abstract}




\section{Introduction}

The epoch of reionization, when the first luminous sources reionized the neutral hydrogen in the intergalactic medium (IGM), is one of the most dramatic periods of evolution in the young universe. 
During this time, ionization fronts impulsively heated reionized gas in the IGM to $\sim 10^4$ K. 
The exact amount of heat injected into the IGM depends on the proprieties of the luminous sources as well as the timing and duration of reionization \citep{mcquinn_2012, davies_2016, D_Aloisio_2019}.
After reionization, the IGM expands and cools through the adiabatic expansion of the universe and inverse Compton scattering off CMB photons. 
The combination of photoionization heating, Compton cooling, and cooling due to the expansion of the universe result in a tight power-law temperature-density relation for most of the IGM gas: 
\begin{equation}
    T = T_0 \Delta^{\gamma - 1}
    \label{eq:temperature-density relation}
\end{equation}
for overdensity $\Delta = \rho / \bar{\rho}$, the mean density of the Universe $\bar{\rho}$, temperature at mean density $T_0$, and an expected slope $\gamma$ \citep{hui_gnedin_1997, puchwein_2015, Mcquinn_upton_2016}. 
However, the low-density IGM has long cooling times, so the thermal memory of reionization can persist for hundreds of Myr such that the thermal state of the IGM just after reionization ends contains important information on the state of the universe during reionization \citep{miralda_escude_1994, hui_gnedin_1997, haehnelt_1998, theuns_2002_wavelet, hui_haiman_2003, lidz_2014, onorbe_2017, Onorbe_2017_planck}.
Describing the thermal state of the IGM ($T_0$ and $\gamma$) just after reionization, $z \sim 5 - 6$, is key to understand the evolution of the universe during reionization. 

The premier probe of the IGM is Ly$\alpha$ absorption along sightlines to bright quasars at high redshift, known as the Ly$\alpha$ forest \citep{gunn_peterson_1965, lynds_1971}. 
The properties of these absorption features are sensitive to the thermal state of the IGM from two effects: Doppler broadening due to thermal motions and Jeans (pressure) smoothing of the underlying baryon distribution.
The rate at which pressure forces erase gravitational fluctuations is set by the local sound speed, and at IGM densities the pressure scale sound crossing time is approximately the Hubble time. 
Therefore, the pressure smoothing scale provides an integrated record of the thermal history of
the IGM \citep{gnedin_hui_1998, kulkarni_2015, Nasir_2016, onorbe_2017, Onorbe_2017_planck, Rorai_2017}.
Both of these effects reduce the small-scale structure of the Ly$\alpha$ forest. 

Several statistics have been used to measure the thermal state of the IGM, including 
the flux probability density \citep{becker_2007, bolton_2008, viel_2009, calura_2012, Lee_2015}, the curvature \citep{becker_2011, boera_2014, gaikwad_2020}, the Doppler parameter distribution \citep{schaye_1999, schaye_2000, ricotti_2000, bryan_2000, mcdonald_2001, rudie_2012, bolton_2010, bolton_2012, bolton_2014, rorai_2018, gaikwad_2020}, and the joint distribution of the Doppler parameters with the Hydrogen Column Density \citep{Hiss_2018}.
One of the most commonly used statistics used to measure the structure of the Ly$\alpha$ forest is the 1D flux power spectrum ($P_{\rm F}(k)$) \citep{theuns_2000, Zaldarriaga_2001, yeche_2017, Walther_2017,Boera_2019, gaikwad_2020}. 
The reduction in small-scale structure in the Ly$\alpha$ forest leads to a cut-off in the power at high $k$ values. 
However, with measurements of higher-redshift quasars, closer to reionization, the optical depth and its scatter for Ly$\alpha$ photons increase \citep{fan_2006, becker_2015}, leading to more absorption and Gunn-Peterson troughs in the Ly$\alpha$ forest. 
Calculating the 1D flux power spectrum at these high redshifts thus mixes high signal-to-noise ratio transmission spikes with noisy absorption troughs, potentially leading to a loss of information. 

Wavelet analysis provides an alternative statistical method to measure the structure of the Ly$\alpha$ forest over a range of characteristic scales \citep{Lidz_2010, garzilli_2012, gaikwad_2020} (though see also \citet{theuns_zaroubi_2000, theuns_2002_wavelet, Zaldarriaga_2002, meiksin_2000}). 
Wavelets are localized in both frequency and real space, which allows them to encode Fourier information while remaining in configuration space. 
Therefore, wavelet analysis has the benefit of keeping the absorption troughs distinct from the transmission spikes because it produces a full decomposition of wavelet amplitudes along the spectrum.
The ultimate statistic used in wavelet analysis is the full wavelet amplitude probability density function (PDF).
The PDF potentially contains more information than the average, which is effectively encoded in the power spectrum. 
However, these wavelet amplitude PDFs are complicated owing to the large correlations between bins in one wavelet amplitude PDF as well as between different wavelet amplitude PDFs. 

Our work builds off and improves upon the previous implementation of wavelet analysis done by \citet{Lidz_2010} and \citet{gaikwad_2020}. 
The work done in \citet{Lidz_2010} used one of the two characteristic wavelet scales explored to constrain the thermal state of the IGM.
Each wavelet scale picks out a frequency in the flux so, to compare to the constraints on the thermal state of the IGM from $P_{\rm F}(k)$, the number of smoothing scales used in wavelet analysis should be comparable to the number of band powers in $P_{\rm F}(k)$.  
Only using one scale will reduce the constraining power of the wavelet amplitude PDFs because it is missing information in other Fourier modes. 
\citet{Lidz_2010} also ignored correlations between the bins in the wavelet amplitude PDFs, potentially significantly affecting the resulting error bars. 
\citet{gaikwad_2020} used eight wavelet scales in their analysis and included the correlations between the bins within each wavelet amplitude PDF. 
Their method still ignored the correlations between the bins for wavelet amplitude PDFs of different scales, again potentially effecting the resulting error bars. 
They also combined their $P_{\rm F}(k)$ measurements with their wavelet PDF measurements that were calculated from the same data set (along with the Doppler parameter distribution and the curvature statistic), ignoring correlations between all these statistics, to get a more precise measurement. 

Our work quantifies the precision of parameter inference using wavelet amplitude PDFs and $P_{\rm F}(k)$.
We show that measuring $T_0$ from our simple thermal model from the wavelet amplitude PDFs results in a 7\% reduction of the 1$\sigma$ errors when compared to the measurement from $P_{\rm F}(k)$ on the same mock data set. 
This confirms the potential that wavelet amplitude PDFs have to improve upon existing constraints on the thermal state of the IGM. 
Our wavelet analysis method uses more scales than previous works and spans the full range of scales probed by $P_{\rm F}(k)$. 
For the first time, we calculate and present the full correlation matrices between the bins of the wavelet amplitude PDFs as well as the cross-correlations between $P_{\rm F}(k)$ and the wavelet amplitude PDFs. 
We also combined the wavelet amplitude PDFs with $P_{\rm F}(k)$ while taking the cross-correlations into account and found that this did not further improve the measurement.
Finally, we characterized the effects of ignoring cross-correlations for the wavelet amplitude PDFs and the combination of the two statistics.

In addition to the thermal state of the IGM, the small-scale structure of the Ly$\alpha$ forest is also sensitive to departures from cold dark matter (CDM), including models of warm dark matter (WDM).
For WDM, the linear power spectrum is exponentially suppressed when compared to CDM on scales smaller than the free-streaming length of the WDM particle \citep{Narayanan_2000}.
The mass of the WDM particle, $m_{\text{WDM}}$, can then be constrained by requiring the initial conditions to have sufficient small-scale power to reproduce the properties of the Ly$\alpha$ forest \citep{Viel_2013, Irsic_2017, garzilli_2017}. 
Wavelet analysis thus also has the potential to improve constraints on the mass of a WDM particle from the small-scale structure in the Ly$\alpha$ forest. 

The structure of this paper is as follows. We describe our procedure for generating simulated Ly$\alpha$ forest sightlines in Section \ref{section:simulation}. 
We then introduce and explore the properties of our wavelet analysis in Section \ref{section:wavelets}.
Our method for statistical inference is laid out in Section \ref{section:stat inference}. 
Our results comparing the measurements from the wavelet analysis and power spectrum is in Section \ref{section:results}. 
We summarize in Section \ref{section:conclusion}.


\section{Simulating Lyman-alpha Forest Spectra} \label{section:simulation}

\subsection{Hydro Simulations}

For this work we use one simulation run that uses the Nyx code. 
Nyx is a cosmological hydrodynamical simulation code designed for simulating the Ly$\alpha$ forest. 
For more details on the numerical methods, scaling, and the heating and cooling rates see \citet{almgren_2013} and \citet{Lukic_2015}. 
We use a standard $\Lambda$CDM cosmological model consistent with the constraints from \citet{Planck_2018}: $\Omega_b = 0.04964$, $\Omega_m = 0.3192$, $\Omega_{\Lambda} = 0.6808$, $h = 0.67038$, $\sigma_8 = 0.826$, and $n_s = 0.9655$. 
The simulation we used has a box size of length, $L_{\rm box} = \SI{20}{\mega\parsec\per\h}$ and $1024^3$ resolution elements. 
To simulate reionization, we use the \textit{flash} model from \citet{Onorbe_2019} which reionizes at $z_{\rm reion}=7.75$, and uses $\Delta T=2 \times 10^4$ to parameterize the instantaneous heat injection from reionization. 
In this framework every cell in the simulation will be ionized at $z=7.75$ and heated to $\Delta T$,  unless the cell was previously ionized by a different process (i.e. collisional ionization). 
We consider two snapshots from this simulation at $z=5$ and $z=6$.
We output 10,000 skewers of the Ly$\alpha$ forest from each snapshot to use in our analysis, which is equivalent of a total pathlength of $\SI{200}{\giga\parsec\per\h}$. 
The pixel scale of the simulation snapshot is $\Delta v = \SI{2.7}{\kilo\meter\per\second}$ at $z=5$ and is $\Delta v = \SI{2.9}{\kilo\meter\per\second}$ at $z=6$.
Since there is a larger dataset available at $z=5$, we focus our work at this redshift.
Figures shown in the main text will be at $z=5$ unless otherwise specified while figures for $z = 6$ are available in Appendix \ref{sec:redshift_6}.

\subsection{Thermal Models}

In the \textit{flash} reionization model, the majority of the IGM follows the tight temperature-density relation of equation \eqref{eq:temperature-density relation} after reionization, see \citet{Onorbe_2019} for more details. 
In order to create simulation Ly$\alpha$ absorption sightlines with different values of $T_0$, we adopt a semi-numerical approach to `paint' on the temperature. 
We do this to each simulation cell using the density output from the simulation and setting the temperature according to equation \eqref{eq:temperature-density relation} with our desired $T_0$. 
This is done for all densities with no cutoff.
This is a simplistic model that does not take into account the full evolution of the thermal state of the IGM. 
However, the purpose of this paper is to present our statistical method and demonstrate its accuracy and precision on simulated data so a simple temperature model for the IGM thermal state is sufficient to achieve these aims. 
We use $\gamma = 1.35$, which was calculated by fitting the initial simulation snapshot to a power law. 
Our thermal grid consists of 81 values of $T_0$ from $\log(T_0) = 3.4$ to $\log(T_0) = 4.4$ with $\Delta \log(T_0) = 0.0125$.
The Ly$\alpha$ opacity, $\tau_{\rm Ly\alpha}$ is related to the temperature via
$\tau_{\rm Ly\alpha} = n_{\rm HI} \sigma_{{\rm Ly}\alpha} \propto  T^{-0.7} / \Gamma_{\rm HI}$, see \citet{Rauch_1998}. 
Because UV background photoionization, $\Gamma_{\rm HI}$, is sourced by complex galaxy physics, it is not uniquely determined by the simulation. 
We therefore follow standard practice and adjust each model to have the same mean flux by rescaling $\tau$ such that $\langle e^{-\tau} \rangle = \langle F \rangle = 0.16$ at $z=5$, which is within $1\sigma$ of the measurement presented in \citet{Boera_2019}.
At $z=6$ we use $\langle F \rangle = 0.011$ which is also consistent with recent measurements \citep{becker_2015, D_aloisio_2018}.

\subsection{Forward Modeling Real Observations} \label{section:observational data}

To mimic realistic observational data from echelle spectrographs, (e.g. from Keck/HIRES, VLT/UVES, and Magellan/MIKE) we forward model a resolution of $R = 30,000$ and a signal to noise ratio per pixel (SNR) of the unabsorbed continuum of 10 (35) at $z=5$ ($z=6$). 
The resolution smooths the flux by a Gaussian filter with $\text{FWHM} = \SI{10}{\kilo\meter\per\second}$ which means our simulations have $\sim 4$ pixels per FWHM of this resolution filter. 
For simplicity, we add flux-independent noise in the following way.
We generate one 10,000 skewer x 1024 length realization of random noise all drawn from a Gaussian with $\sigma_N = 1 / \text{SNR}$ and add this noise realization to every temperature model. 
An example skewer of our initial and forward-modeled data is shown in the top panels of Figures \ref{fig:amps_skewer} and \ref{fig:noisy_amps_skewer} respectively.
Using the same noise realizations over the different models ensures that different noise realizations will not adversely affect the inference on the $T_0$ for mock data. 

We assume a fiducial data set size of 8 quasar spectra at both $z = 5$ and $z=6$ that probe a redshift interval of $\Delta z = 0.2$ per quasar for a total pathlength of $\Delta z = 1.6$ (equivalent to 29 skewers). 
In the discussion going forward, each mock data set consists of a random selection of 29 skewers without replacement. 


\section{Wavelet Analysis} \label{section:wavelets}

\subsection{Formalism} \label{section:wavelet formal}

\begin{figure}
	\includegraphics[width=\columnwidth]{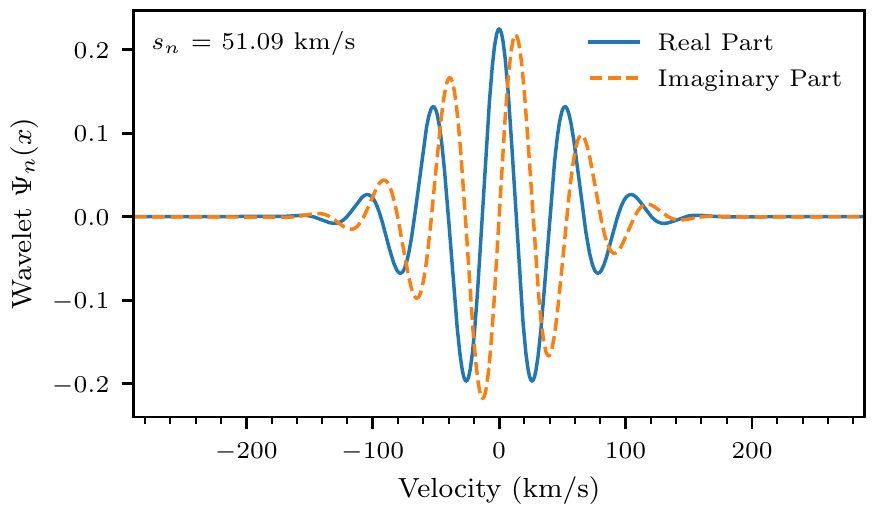}
    \caption{A complex Morlet Wavelet filter in real space with $s_n = \SI{51.09}{\kilo\meter\per \second}$. The solid line shows the real part of the wavelet while the dashed line shows the imaginary part. The width of the oscillations are set by the smoothing scale.}
    \label{fig:wavelet}
\end{figure}

Wavelets are localized in frequency and configuration space which allows wavelet amplitudes provide a breakdown of Fourier information at all locations along a quasar sightline. 
Following \citet{Lidz_2010}, we calculate wavelet amplitudes from a "complex Morlet wavelet", which is shown in Figure \ref{fig:wavelet} and has the functional form:
\begin{equation}
\centering
    \Psi_n(x) = A \exp(i k_0 x) \exp \left[ - \frac{x^2}{2 s_n^2} \right].
\end{equation}
The normalization, $A$, is set by requiring that $\abs{\Psi_n(k)} = 1$. With this normalization, the Fourier transform of a complex Morlet wavelet is 
\begin{equation}
    \Psi_n(k) = \pi^{-1/4} \sqrt{\frac{s_n}{\Delta u}} \exp \left[ - \frac{(k-k_0)^2 s_n^2}{2}  \right].
\end{equation}
This is a Gaussian in configuration space centered on $k_0$ with width $\sigma_k = \sqrt{2}/s_n$. We also require that  $k_0 s_n = 6$ to ensure these filters have a close to zero mean. 

To begin the analysis on our simulated spectra, we first compute the flux contrast of the Ly$\alpha$ forest, $\delta_F$:
\begin{equation}
    \delta_F = \frac{F - \bar{F}}{\bar{F}}.
    \label{eq:delta_flux}
\end{equation}
Then we convolve this flux contrast field with a wavelet filter of smoothing scale $s_n$ resulting in a filtered spectrum, $a_n$:
\begin{equation}
    a_n(x) = \int dx' \Psi_n(x - x')\delta_F(x') 
\end{equation}
The filtered spectrum is a complex number, the modulus of which is called the "wavelet amplitude" $A_n(x)=\abs{a_n(x)}^2.$ 
We define the power spectrum as
\begin{equation}
    \langle \delta_F(k) \delta_F(k^\prime) \rangle = 2 \pi P_F(k) \delta_D(k - k^\prime)
\end{equation}
where $\delta_D$ is the Dirac Delta function. With this definition of the power, the average wavelet amplitude is
\begin{equation}
    \langle A_n(x) \rangle = \int^{\infty}_{-\infty} dk' 2\pi [\Psi_n(k') ]^2 P_F(k').
    \label{eq:power wavelet equivalence}
\end{equation}
In words, this means that the average wavelet amplitude is the power spectrum averaged over a Gaussian 
centered on wave-number $k_0 = 6/s_n$ with standard deviation $\sqrt{2}/s_{n}$. 
Therefore, this average wavelet amplitude is effectively a band-power.

\begin{figure*}
	\includegraphics[width=\columnwidth*2]{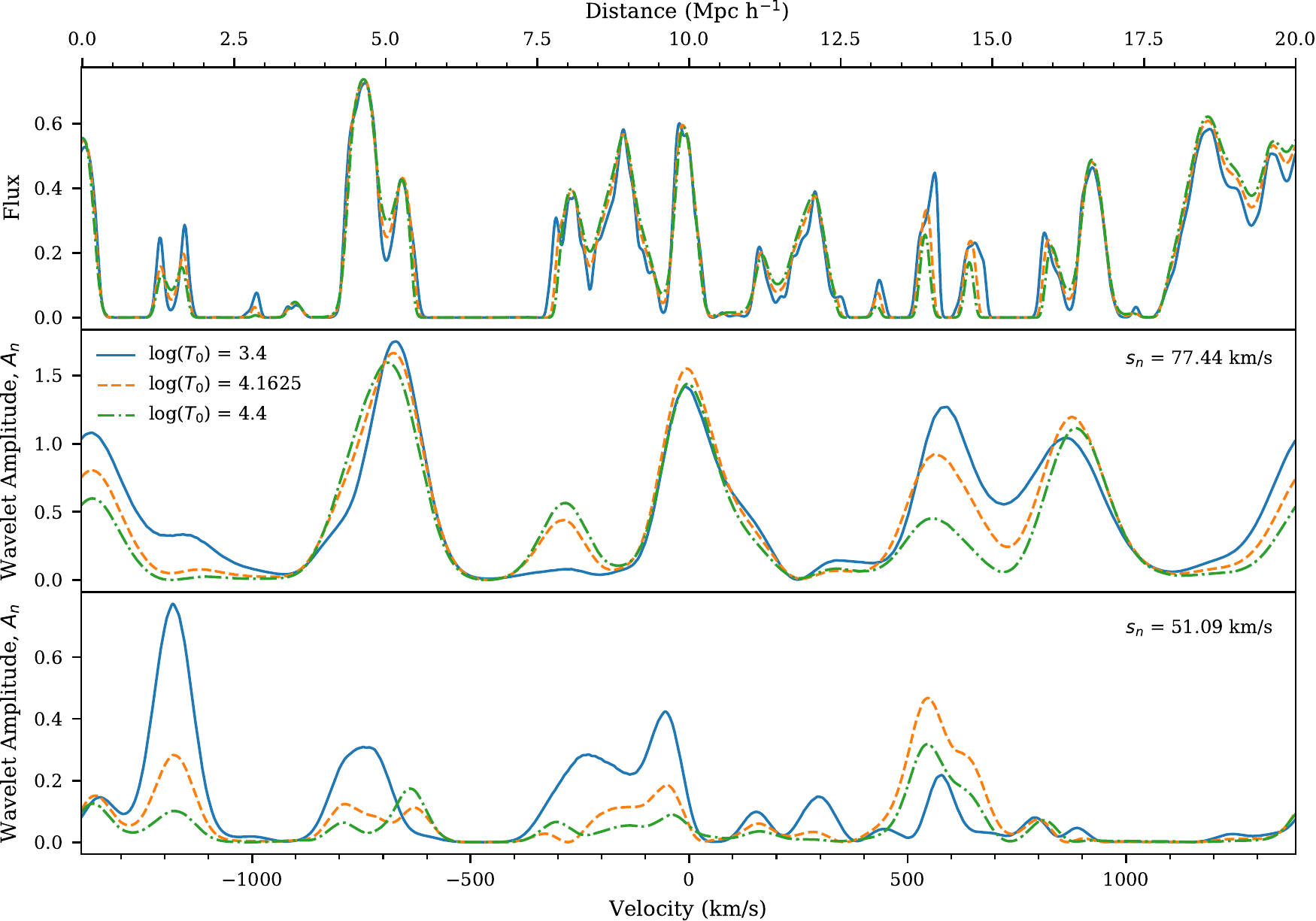}
    \caption{
    The top panel shows the flux from one simulation skewer at $z=5$ for the three different values of $T_0$: $\log(T_0) = 3.4$ (blue), $\log(T_0) = 4.1625$ (orange), and $\log(T_0) = 4.4$ (green).
    The middle panel shows the wavelet amplitude spectra for $s_n = \SI{77.44}{\kilo\meter\per\second}$ and the bottom show the wavelet amplitude spectra for $s_n = \SI{51.09}{\kilo\meter\per\second}$, both with the same $T_0$ values as the top panel. 
    This shows that the largest values of wavelet amplitudes correspond to peaks in the flux that are roughly the same width as the oscillations set by the wavelet smoothing $s_n$ scale.
    }
    \label{fig:amps_skewer}
\end{figure*}

Two wavelet amplitude spectra for an ideal simulated skewer at $z=5$ are shown in the bottom two panels of Figure \ref{fig:amps_skewer}. 
For illustrative purposes in this section, we will mainly show wavelet amplitudes for $s_n = \SI{51.09}{\kilo\meter\per \second}$ though Figure \ref{fig:amps_skewer} also shows $s_n = \SI{77.44}{\kilo\meter\per\second}$ for a comparison. 
Ultimately in our analysis at $z=5$, we will use fifteen logarithmically spaced values of $\SI{2200}{\kilo\meter\per\second} > s_n > \SI{5}{\kilo\meter\per\second}$ as described in Section \ref{section:amp likelihood}. 
For $z=6$ we still use fifteen logarithmically spaced values of $s_n$ with slightly shifted values due to the redshift dependence of the simulation resolution and box size. 

The purpose of Figure \ref{fig:amps_skewer} is to show the relationship between the flux and wavelet amplitudes for different smoothing scales. 
The top panel shows the flux for the three different values of $T_0$: $\log(T_0) = 3.4$ (blue), $\log(T_0) = 4.1625$ (orange), and $\log(T_0) = 4.4$ (green). 
The middle panel shows the wavelet amplitude spectra for $s_n = \SI{77.44}{\kilo\meter\per\second}$ and the bottom show the wavelet amplitude spectra for $s_n = \SI{51.09}{\kilo\meter\per\second}$, both with the same values of $T_0$ as the top panel. 
The smoothing scale sets the size of the features in the flux that are picked out, when the smoothing scale and the feature size in the flux align the resulting wavelet amplitude is greater. 
The middle panel has a greater value of $s_n$ than the bottom panel, so it is going to pick out wider features in the flux. 
Consider the peak in the flux at $\sim \SI{550}{\kilo\meter\per\second}$, which is smoother (and smaller) for $\log(T_0) = 4.1625$ (orange) flux than for $\log(T_0) = 3.4$ (blue).
The corresponding wavelet amplitudes in the middle panel are greatest for $\log(T_0) = 3.4$ (blue) while the bottom panel are greatest for $\log(T_0) = 4.1625$ (orange), showing that the smaller feature in the flux agreed better with the smaller smoothing scale, as expected. 
The flux at $\log(T_0) = 4.4$ (green) is even smoother than the flux at $\log(T_0) = 4.1625$ (orange) but it does not have greater wavelet amplitude values in the bottom panel, this is because this peak corresponds to an ever smaller smoothing scale. 

Figure \ref{fig:amps_skewer} shows that the largest values of wavelet amplitudes correspond to peaks in the flux that are roughly the same width as the oscillations set by the wavelet smoothing $s_n$ scale (an example of these oscillations can be seen in Figure \ref{fig:wavelet}). 
There can be offsets between features in flux and the corresponding features in the wavelet amplitude spectra, since the wavelets pick out features with the specific width set by $s_n$ and, at larger scales, the wavelet will combine multiple features in the flux spectrum. 
This figure also demonstrates how wavelet analysis presents Fourier information in configuration space since the different wavelet amplitudes values convey frequency information along the quasar sightline. 

\begin{figure}
	\includegraphics[width=\columnwidth]{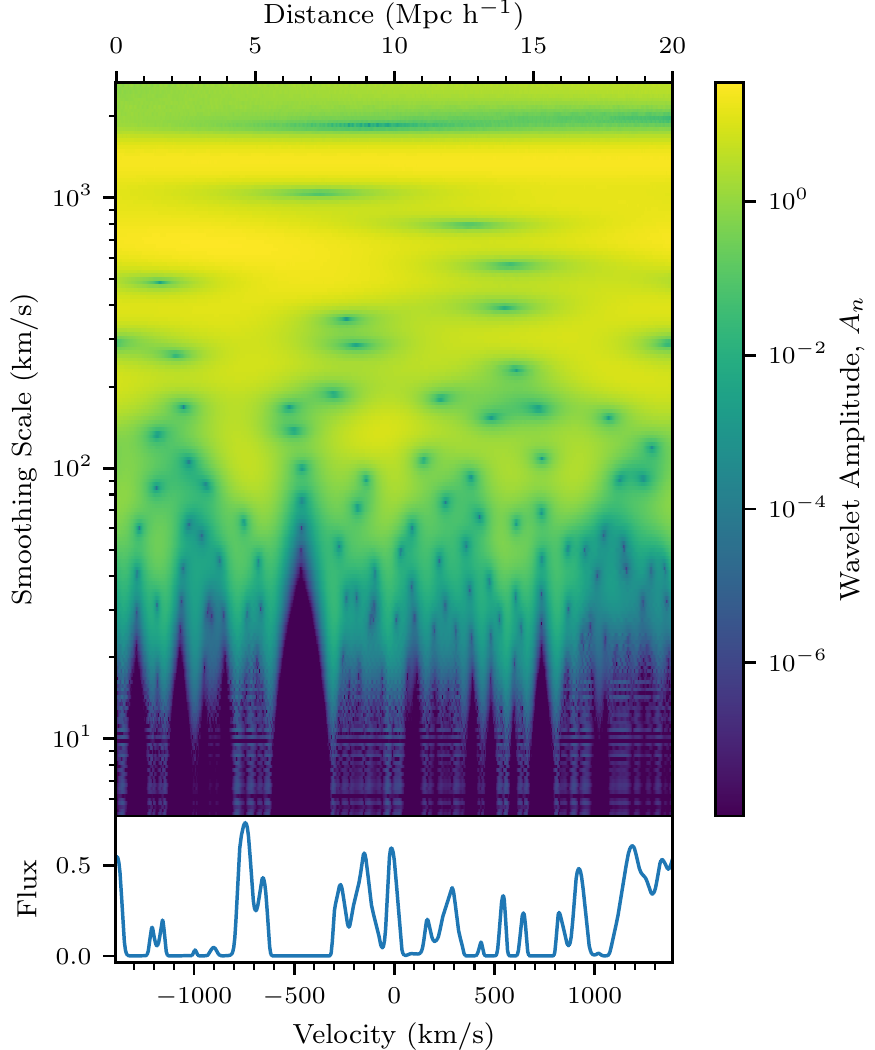}
    \caption{The bottom panel shows the flux from one simulation skewer with $\log(T_0) = 4.1625$. 
    The top panel shows a periodogram of the wavelet amplitudes along the skewer for different smoothing scales. 
    The different values of $s_n$ will set different widths of oscillations that they pick out from the spectrum.
    This plot compares the location of the peaks and troughs in the wavelet amplitudes for different smoothing scales. 
    It shows correlations between troughs at different values of $s_n$, for example the trough in the flux at $\SI{-500}{\kilo\meter\per\second}$ can clearly be seen in the wavelet amplitudes for scales up to $s_n \sim \SI{40}{\kilo\meter\per\second}$. 
    Note that the minimum wavelet amplitude shown on the plot is fixed at $10^{-8}$ for visual purposes.}
    \label{fig:periodogram_clean}
\end{figure}

In order to compare the spatial correlations between different values of $s_n$, consider Figure \ref{fig:periodogram_clean}.
The top panel of this figure shows a color plot of the wavelet amplitudes for different values of $s_n$ along one line of sight; this is known as a "periodogram". 
The bottom panel of the plot is the flux used the calculate the wavelet amplitudes, which is the same as in Figure \ref{fig:amps_skewer} for $\log(T_0) = 4.1625$. 
The large trough in the flux at $\SI{-500}{\kilo\meter\per\second}$ is seen in the wavelet amplitudes for scales up to $s_n \sim \SI{40}{\kilo\meter\per\second}$. 
The other troughs in the flux, such as the one near $\SI{750}{\kilo\meter\per\second}$, are also seen in the wavelet amplitudes across multiple smoothing scales, most prominently at the smaller values of $s_n$. 
The overall decline in the average wavelet amplitude value for smaller values of $s_n$ follows from the cutoff in the power spectrum, as is expected from equation \eqref{eq:power wavelet equivalence}. 

\begin{figure*}
	\includegraphics[width=\columnwidth*2]{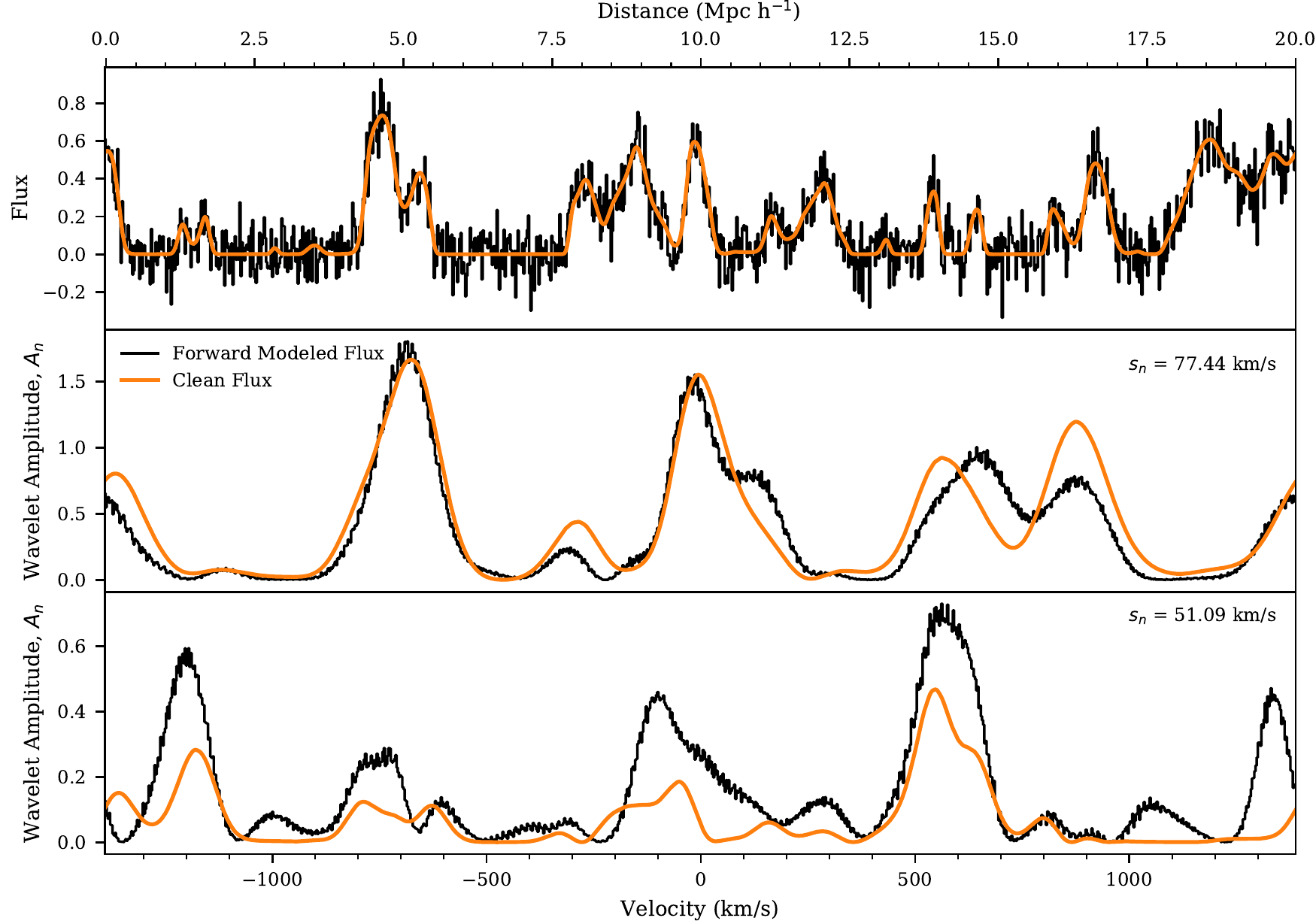}
    \caption{
        The top panel shows the flux for the $\log(T_0) = 4.1625$ model from the simulation (orange line) as well as forward modeled with resolution and noise (black histogram).
        The middle panel shows the wavelet amplitude spectra for $s_n = \SI{77.44}{\kilo\meter\per\second}$ and the bottom show the wavelet amplitude spectra for $s_n = \SI{51.09}{\kilo\meter\per\second}$, both with the same $T_0$ value as the top panel. 
        The simulation skewer is the same as that shown in Figure \ref{fig:amps_skewer}.
        This shows the effect noise has on the flux and the resulting wavelet amplitudes for one skewer. 
    }
    \label{fig:noisy_amps_skewer}
\end{figure*}

As discussed in Section \ref{section:observational data}, we forward modeled our simulation skewers to mimic real data by including the effects of the resolution and noise. 
We illustrate the change in the flux as well as the wavelet amplitudes for one example skewer with $\log(T_0) = 4.1625$ in Figure \ref{fig:noisy_amps_skewer}. 
Note that the ``clean'' flux and wavelet amplitudes in this figure matches the model in Figure \ref{fig:amps_skewer} from the same temperature model.
From the Figure, we see that adding noise to the flux is able to shift features, add additional features, and change the amplitude of features in the wavelet amplitude spectrum.
These effects are more prominent on smaller scales, as seen in the large differences between the models in the bottom panel of Figure \ref{fig:noisy_amps_skewer}, since the noise power becomes comparable or greater than the flux power at these scales. 
Overall, this panel has greater values and more high-valued wavelet amplitudes for the forward modeled skewers than those without noise.

\subsection{Wavelet Probability Density Function}

As illustrated by Figures \ref{fig:amps_skewer}, \ref{fig:periodogram_clean}, and \ref{fig:noisy_amps_skewer} wavelet analysis converts the flux of the Ly$\alpha$ forest into wavelet amplitude spectra parameterized by $s_n$. 
The average value of the wavelet amplitude spectra corresponds to $P_{\rm F}(k)$ via equation \eqref{eq:power wavelet equivalence}. 
The statistic we measure in our analysis is the wavelet amplitude probability density function (PDF), since this contains information on the full distribution of the wavelet amplitude values, rather than only the average.

\begin{figure}
	\includegraphics[width=\columnwidth]{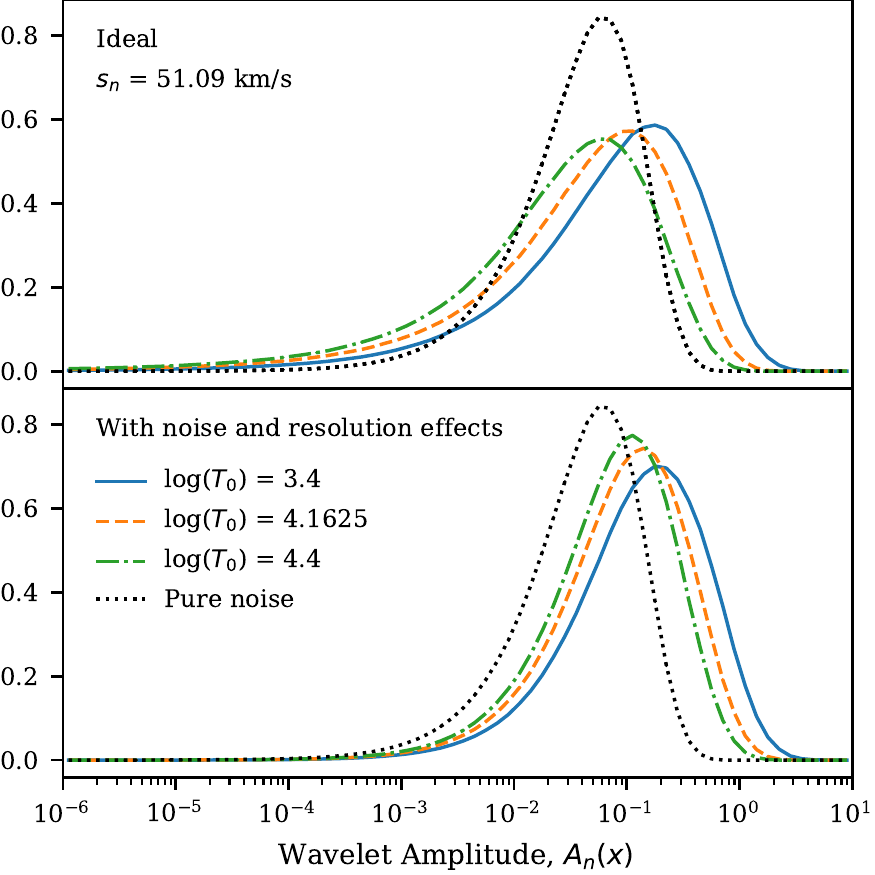}
    \caption{This figure shows the PDFs for $s_n = \SI{51.09}{\kilo\meter\per\second}$ for three different values of $T_0$: $\log(T_0) = 3.4$ (blue), $\log(T_0) = 4.1625$ (orange), and $\log(T_0) = 4.4$ (green). 
    The top panel shows the PDFs calculated from 10,000 ideal simulation skewers (equivalent to a pathlength of $\SI{200}{\giga\parsec\per\h}$). 
    The bottom panel shows the PDFs after forward-modeling with resolution and noise to mimic real data. 
    In both the top and the bottom panel, the black dotted line shows the same PDF for pure noise draws with SNR$ = 10$ and our pixel resolution. 
    The difference in the mean values of these PDFs in each panel is expected from $P_{\rm F}(k)$ and equation \eqref{eq:power wavelet equivalence}.
    The main effect of forward-modeling is the suppression of all wavelet amplitude values below $\sim 10^{-3}$, where the data is beginning to be dominated by the noise. 
    }
    \label{fig:two_panel_amp_pdf}
\end{figure}

PDFs for $s_n = \SI{51.09}{\kilo\meter\per\second}$ are shown in Figure \ref{fig:two_panel_amp_pdf} for three different values of $T_0$: $\log(T_0) = 3.4$ (blue), $\log(T_0) = 4.1625$ (orange), and $\log(T_0) = 4.4$ (green). 
The top panel shows the PDFs calculated from the ideal simulation with clean flux. 
The bottom panel shows the PDFs after forward-modeling the simulation output with resolution and noise to mimic real data, as discussed in Section \ref{section:observational data}. 
In both the top and the bottom panel, the black dotted line shows the same PDF for pure noise draws with SNR$ = 10$ and our pixel resolution. 

In the top panel, the ideal PDFs are skewed to the left, with lower IGM temperatures corresponding to a higher mean value, as is expected from $P_{\rm F}(k)$ and equation \eqref{eq:power wavelet equivalence}. 
The main effect of forward-modeling is the shift of the PDF from small values to larger values as was also seen in the bottom panel of Figure \ref{fig:noisy_amps_skewer}. 
This causes the suppression of wavelet amplitude values below $\sim 10^{-3}$.
Initially the $\log(T_0) = 4.4$ (green) PDF had the largest tail below $10^{-3}$, so the shift from small values to large values causes this model to change most dramatically from the top to the bottom panel. 
The PDF both shifts to the right and greatly increases the value of the PDF at the peak. 
The PDFs on the bottom panel are much more similar to the PDF for pure noise than in the top panel, which shows how the noise PDF is able to dominate over the signal. 
As the smoothing scale decreases and the overall PDF values decrease with it, as is inferred from Figure \ref{fig:periodogram_clean} and equation \eqref{eq:power wavelet equivalence}, the PDFs will become more dominated by the noise contribution. 

Figure \ref{fig:two_panel_amp_pdf} demonstrates the ability of the wavelet amplitude PDF to differentiate $T_0$ models both with and without forward modeling.
This confirms that the wavelet amplitude PDFs are promising statistics to measure the thermal state of the IGM.
In addition, it illustrates how the PDF quantifies the full distribution of wavelet amplitudes for multiple sightlines, rather than the values along one sightline or the average value which is encoded in the the power spectrum.

\section{Statistical Methods} \label{section:stat inference}

The goal of this paper is to calculate the statistical precision with which a realistic quasar data set can constrain the parameters governing the small-scale structure of the IGM, here limited to $T_0$, using wavelet analysis, specifically wavelet amplitude PDFs. 
The precision from this method can then be directly compared to the canonical approach using $P_{\rm F}(k)$. 
We will also consider the precision achieved when combining the wavelet amplitude PDFs and power spectrum as has recently been attempted in the literature \citep{gaikwad_2020}. 

To calculate the statistical precision, we will use Bayes' Theorem:
\begin{equation}
    P(T_0 \vert \text{data}) = \frac{P(\text{data} \vert T_0) P(T_0)}{P(\text{data})}.
    \label{eq:bayes}
\end{equation}
Here the "data" vector depends on the statistical method for which we are calculating the precision. 
For the power spectrum, the "data" are the band-powers comprising $P_{\rm F}(k)$. 
For the wavelet analysis, we will use multiple values of $s_n$ and thus have multiple wavelet amplitude PDFs we must consider. 
In this case, the "data" will be the wavelet amplitude PDFs concatenated one after the other from largest to smallest $s_n$ (which corresponds to smallest to largest $k$). 
Finally, when combining the wavelet and power spectrum analysis, the "data" vector will be the concatenated PDFs vector from the wavelet case with $P_{\rm F}(k)$ added onto the end of it. 

We assume a flat prior, $P(T_0)$, over the range of $T_0$ values we have simulation data for and will normalize the posterior, $P(T_0 \vert \text{data})$, to unity so we don't need to explicitly calculate $P(\text{data})$. 
In order to calculate the likelihood, $P(\text{data} \vert T_0) = \mathcal{L}$, we assume a multivariate Gaussian distribution. 
This likelihood has the form:
\begin{equation}
        \mathcal{L} =  \frac{1}{\sqrt{\det(\Sigma) (2 \pi)^{n}}} \exp \left( -\frac{1}{2} (\text{data} - \text{model})^{\text{T}} \Sigma^{-1} (\text{data} - \text{model}) \right) 
        \label{eq:gauss_like}
\end{equation}
where $\Sigma = \Sigma(T_0)$ is the model dependent covariance matrix, $n$ is the number of points in the data vector. 
Both the data and model vectors depend on the statistic we are using and will be discussed in their respective sections. 
The choice of a multivariate Gaussian distribution for the likelihood has been used in previous wavelet studies \citep{Lidz_2010, gaikwad_2020} as well as for studies using the Ly$\alpha$ forest flux PDF \citep{Lidz_2006, Eilers_2017}. 
The base assumption is that each band power for $P_F(k)$  or each bin of the wavelet amplitude PDFs are Gaussian distributed. We show that this assumption is valid for our data in Appendix \ref{section:appendix likelihood choice}.

In our analysis, we estimate the covariance matrix from mock draws of the data by
\begin{equation}
    \Sigma(T_0) = \frac{1}{N_{\text{mocks}}} \sum_{i=1}^{N_{\text{mocks}}}(\text{mock}_i - \text{model})(\text{mock}_i - \text{model})^{\text{T}}
    \label{eq:covariance}
\end{equation}
where $N_{\text{mocks}}$ is the number of forward-modeled mock draws used.
This method estimates a model dependent covariance, not the covariance of the data itself, since we are using many draws in our calculation.
For the power spectrum calculation we use $N_{\text{mocks}} = 5,000$.
We increase the number of mocks to $N_{\text{mocks}} = 1,000,000$ for the wavelet amplitude PDFs and the combination of the power spectrum and the wavelet amplitude PDFs, since these matrices are much larger with more values close to zero. 
Note that mocks are a random combination of 29 skewers without replacement. 
In theory, there are $(10,000!)/(29! \times 9,971!) \approx 10^{85}$ unique sets of 29 skewers from 10,000 skewers.
This means that mock data sets will be correlated since they will contain skewers that are also in other mock data sets. 
However, we do not approach the the total possible number of combinations for these calculations and expect this effect to be negligible. 

To visualize the covariance matrix for each method, we define the correlation matrix, $C$. The correlation matrix is the covariance matrix with the diagonal normalized to 1. 
This is done to the $i$th, $j$th element by
\begin{equation}
        C_{ij} = \frac{\Sigma_{ij}}{\sqrt{\Sigma_{ii}\Sigma_{jj}}}.
        \label{eq:correlation}
\end{equation}

\subsection{Power Spectrum Likelihood} \label{section:power likelihood}

\begin{figure}
	\includegraphics[width=\columnwidth]{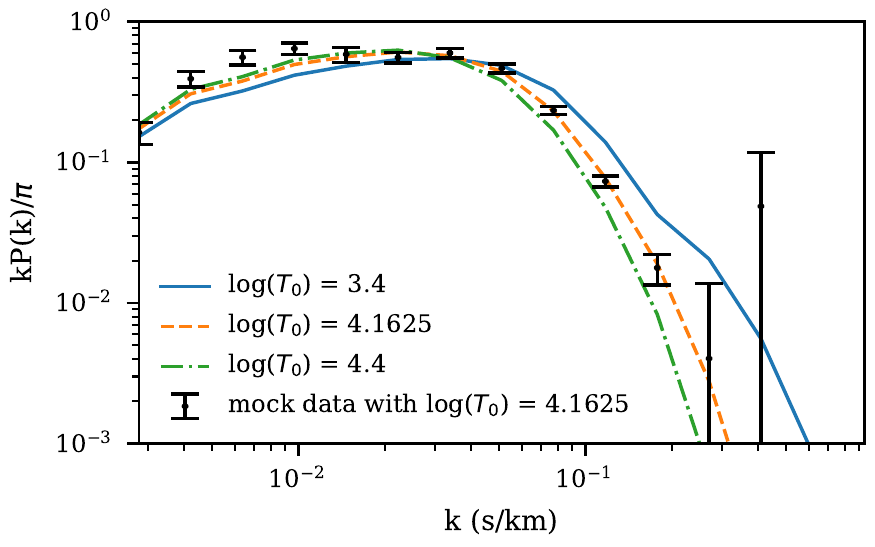}
    \caption{The power spectrum measurement, $P_{\rm F}(k)$, for one mock data set with $\log(T_0) = 4.1625$ (black points). 
    The $1\sigma$ error bars are calculated from the square root of the diagonal of the covariance matrix.
    Also shown are model values of the power spectra for three different values of $T_0$: $\log(T_0) = 3.4$ (blue), $\log(T_0) = 4.1625$ (orange), and $\log(T_0) = 4.4$ (green). 
    }
    \label{fig:one_mock_power}
\end{figure}

The resolution modifies $P_{\rm F}(k)$ in a known way that can be corrected via the Fourier transform of the Gaussian resolution filter. 
There is also a white noise contribution to $P_{\rm F}(k)$ due to the spectral noise which can be subtracted off. 
Therefore, the well known estimator for the true power is: 
\begin{equation}
    P_{\text{true}}(k) = \Bigg\langle \frac{P_{\text{raw}}(k) - P_{\text{noise}}(k)}{W^2_R(k, \sigma_R, \Delta v)} \Bigg\rangle
    \label{eq:power correction}
\end{equation}
where $W_R(k, \sigma_R, \Delta v)$ is the Window function
\begin{equation}
    W_R (k, \sigma_R, \Delta v) = \exp \left( - \frac{1}{2} (k \sigma_R)^2 \right) \frac{\sin(k\Delta v / 2)}{(k \Delta v / 2)}.
    \label{eq:window function}
\end{equation}
We have Gaussian white noise with $\text{SNR} = 10$ added to the flux contrast, adding an extra factor of $\bar{F}$. The noise power is flat and has a value of 
\begin{equation}
    P_{\text{noise}}(k) = \Delta v \left( \frac{1}{\text{SNR} \cdot \bar{F}} \right)^2  
    \label{eq:noise power}
\end{equation}
where the factor of $\Delta v$ is our velocity pixel grid spacing which is $\Delta v = \SI{2.7}{\kilo\meter\per\second}$ at $z=5$. 

For $R = 30,000$, $\exp \left( - \frac{1}{2} (k \sigma_R)^2 \right) < 0.24$ when $k \geq 0.4$. 
This implies that $W^2_R(k, \sigma_R, \Delta v) < .06$ when $k \geq 0.4$, so correcting these band-powers by this window function means dividing a noisy quantity, $P_F(k)$, by a very small number. 
When correcting by the window function, these band-powers blow up and the model covariance matrices we calculate via equation \eqref{eq:covariance} are singular and ill-posed for inversion.  
We therefore choose to not correct by $W_R(k)$ in the "model" and "mock" data to ensure well-behaved covariance matrices. 
However, for visualization purposes we always show the resolution corrected power of equation \eqref{eq:power correction} in the figures. 
The "model" is the power calculated from 10,000 flux skewers forward modeled with the resolution but not the noise, since there is no need to add additional noise when computing the mean.  
10,000 skewers is equivalent to a total pathlength of $\SI{200}{\giga\parsec\per\h}$.
We calculate the "mock" data by computing the average $P_{\rm F}(k)$ for 29 fully forward modeled skewers and then subtracting off the noise power, equation \eqref{eq:noise power}. 
This data set size is equivalent to an 8 quasar data set as discussed in Section \ref{section:observational data}. 

To choose the $k$ values for our mock, we used 15 logarithmic band-powers spanning from $2 \pi / l_{\rm skewer} = \SI{0.0023}{\second\per\kilo\meter}$ to $\pi / \Delta v = \SI{1.2}{\second\per\kilo\meter}$ at $z=5$.
The centers of the band-powers are listed in the first column of Table \ref{tab:limit table}. 
We chose 15 band-powers in order to fully sample the shape of the power spectra while ensuring the low $k$ (large scales) band-powers were populated by the discrete Fourier transform of the data.
Figure \ref{fig:one_mock_power} shows the power spectrum measurement, $P_{\rm F}(k)$, for one mock data set with $\log(T_0) = 4.1625$ (black points) at $z=6$. 
An equivalent figure for $z=6$ can be found in Appendix \ref{fig:z6_power_mock}.
The $1\sigma$ error bars are calculated from the square root of the diagonal of the covariance matrix.
Also shown are three models of the power spectra for three different values of $T_0$: $\log(T_0) = 3.4$ (blue), $\log(T_0) = 4.1625$ (orange), and $\log(T_0) = 4.4$ (green).
This mock data set visually seems to best agree with the model for $\log(T_0) = 4.1625$ (orange) for $k > 0.5$, which is the true $T_0$ of the model. 

\begin{figure}
	\includegraphics[width=\columnwidth]{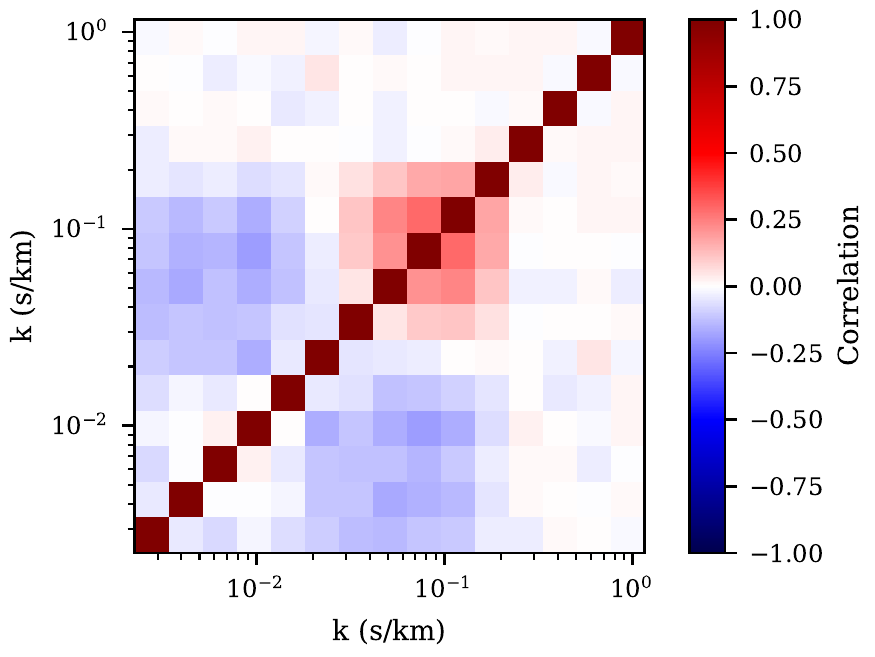}
    \caption{
        The correlation matrix for the power spectrum at $\log(T_0) = 4.1625$.
        The positive (red) and negative (blue) correlations on scales $k < \SI{0.2}{\second\per\kilo\meter}$ arise from underlying spatial correlations of the Ly$\alpha$ forest. 
        The very weak correlations seen in the regions where $k > \SI{0.2}{\second\per\kilo\meter}$ are due to uncorrelated random Gaussian noise which dominates the signal on small-scales (high $k$). 
    }
    \label{fig:power_correlation}
\end{figure}

The model correlation matrix (see equation \eqref{eq:correlation}) for the power spectrum at $\log(T_0) = 4.1625$ is shown in Figure \ref{fig:power_correlation}. 
There are positive correlations (red) between the band-powers where $4 \times 10^{-2} \SI{}{\second\per\kilo\meter} \lesssim k \lesssim \SI{0.2}{\second\per\kilo\meter}$. 
The correlations between band-powers with $4 \times 10^{-2} \SI{}{\second\per\kilo\meter} \lesssim k \lesssim \SI{0.2}{\second\per\kilo\meter}$ and $k < 4 \times 10^{-2} \SI{}{\second\per\kilo\meter}$ the correlations are negative (blue). 
This behavior arises from the underlying spatial correlations of the Ly$\alpha$ forest and is consistent with what has been seen for real data \citep{Walther_2017, Boera_2019}.
At values of $k > 0.2 \SI{}{\second\per\kilo\meter}$ (the smallest scales) noise dominates over the power spectrum, since the Ly$\alpha$ forest power spectrum exhibits a thermal cut-off at high-$k$, whereas the noise power spectrum is flat. 
Random Gaussian noise is uncorrelated, making it hard to recover the signal from the Ly$\alpha$ forest and results in the very weak correlations shown in the correlation matrix in the regions where $k > \SI{0.2}{\second\per\kilo\meter}$. 
We looked into the correlation matrix for SNR $= 50$ and SNR $= 100$ and found the values of the correlation matrix in the column above $k = \SI{0.178}{\second\per\kilo\meter}$ were stronger. 
This agrees with our interpretation of the the weak correlations in Figure \ref{fig:power_correlation} since with higher SNR the noise power is smaller and will not dominate until higher $k$.

\begin{table}
    \centering
    \caption{The first column contains the band-powers for the power spectrum. 
    Next are the corresponding smoothing scales ($s_n$) used to calculate the wavelet amplitudes. 
    The last two columns contain the minimum and maximum values used for the PDF estimation for each smoothing scale. 
    These were chosen as the 1.5th and 98.5th percentiles of the data for the whole thermal grid at these smoothing scales.
    }
	\label{tab:limit table}
    \begin{tabular}{|c|c|c|c|}
        \hline
        $k$ (\SI{}{\second\per\kilo\meter}) & $s_n = 6/k$ (\SI{}{\kilo\meter\per\second}) & min $\log(A_n)$ & max $\log(A_n)$ \\ \hline
        0.00278 & 2157.35 & -0.671 & 1.683  \\ 
        0.00422 & 1423.32 & -0.499 & 1.826  \\ 
        0.00639 & 939.04  & -0.789 & 1.643  \\ 
        0.00968 & 619.54  & -0.876 & 1.574  \\ 
        0.0147  & 408.74  & -1.026 & 1.474  \\ 
        0.0222  & 269.67  & -1.241 & 1.344  \\ 
        0.0337  & 177.91  & -1.508 & 1.173  \\ 
        0.0511  & 117.38  & -1.863 & 0.931  \\ 
        0.0775  & 77.44   & -2.256 & 0.564  \\ 
        0.117   & 51.09   & -2.651 & 0.039  \\ 
        0.178   & 33.71   & -2.943 & -0.430 \\ 
        0.270   & 22.24   & -3.032 & -0.585 \\ 
        0.409   & 14.67   & -3.073 & -0.627 \\ 
        0.620   & 9.68    & -3.129 & -0.688 \\ 
        0.939   & 6.39    & -3.280 & -0.830 \\ \hline
    \end{tabular}
\end{table}

\subsection{Wavelet Amplitude PDF Likelihood} \label{section:amp likelihood}

In previous work, \citet{Lidz_2010} measured the thermal state of the IGM with wavelets by assuming a Gaussian likelihood and ignoring correlations between PDF bins as well as between PDFs from different smoothing scales. 
\citet{gaikwad_2020} measured the thermal state with wavelets assuming a Gaussian likelihood including correlations between bins of the same PDF but not between the PDFs for different smoothing scales. 
Here, we improve upon these previous work and present the likelihood calculation taking into consideration all correlations, both between PDF bins and between PDFs of different smoothing scales. 

For wavelet amplitude PDFs, there is no analytic way to correct for the window function and subtract off the full noise PDF. 
Instead, we choose to use 10,000 forward-modeled skewers (with resolution and noise) to calculate the "model" wavelet amplitude PDFs, which is equivalent to calculating the wavelet amplitude PDFs for a total pathlength of $\SI{200}{\giga\parsec\per\h}$.
The "mock" data is calculated from the same forward-modeled skewers as the "model", though "mock" data is the average of 29 skewers (equivalent to an 8 quasar data set, see Section \ref{section:observational data}). 

As was mentioned in Section \ref{section:wavelet formal}, we use fifteen values of $s_n$ to get fifteen wavelet amplitude PDFs. 
These fifteen $s_n$ correspond to the centers of the power spectrum band-powers, $k$, that were discussed in Section \ref{section:power likelihood} and are listed in the second column of Table \ref{tab:limit table}. 
Qualitatively, this should ensure that the wavelet amplitude PDFs contain at least as much information as the power spectrum due to equation \eqref{eq:power wavelet equivalence}, allowing us to compare the resulting precision on an equal footing. 
To estimate the wavelet amplitude PDFs for each smoothing scale, we calculate histograms. 
This introduces three histogram parameters into our analysis: the maximum wavelet amplitude considered, the minimum wavelet amplitude considered, and the number of bins in the histogram. 

When selecting the minimum and maximum wavelet amplitude considered for our PDF estimation, we want to ensure that all bins will be populated for the whole thermal grid so that the covariance matrix is well posed for inverting. 
We also want to make sure the maximum and minimum values span a large enough range to capture the most significant differences in the shape of the PDF. 
For these reasons, we chose the maximum and minimum values for the PDFs by calculating the 1.5th and 98.5th percentile of the "model" wavelet amplitudes calculated for every thermal model in our grid. 
The maximum and minimum values of the wavelet amplitudes considered for each smoothing scale $s_n$ are listed in Table \ref{tab:limit table}. 
We also need to select a number of bins that will sufficiently sample the shape of the PDF without making the data vector too long and the covariance matrix ill-suited for inversion. 
We found that 10 bins was a reasonable choice to achieve these aims. 

\begin{figure*}
	\includegraphics[width=\columnwidth*2]{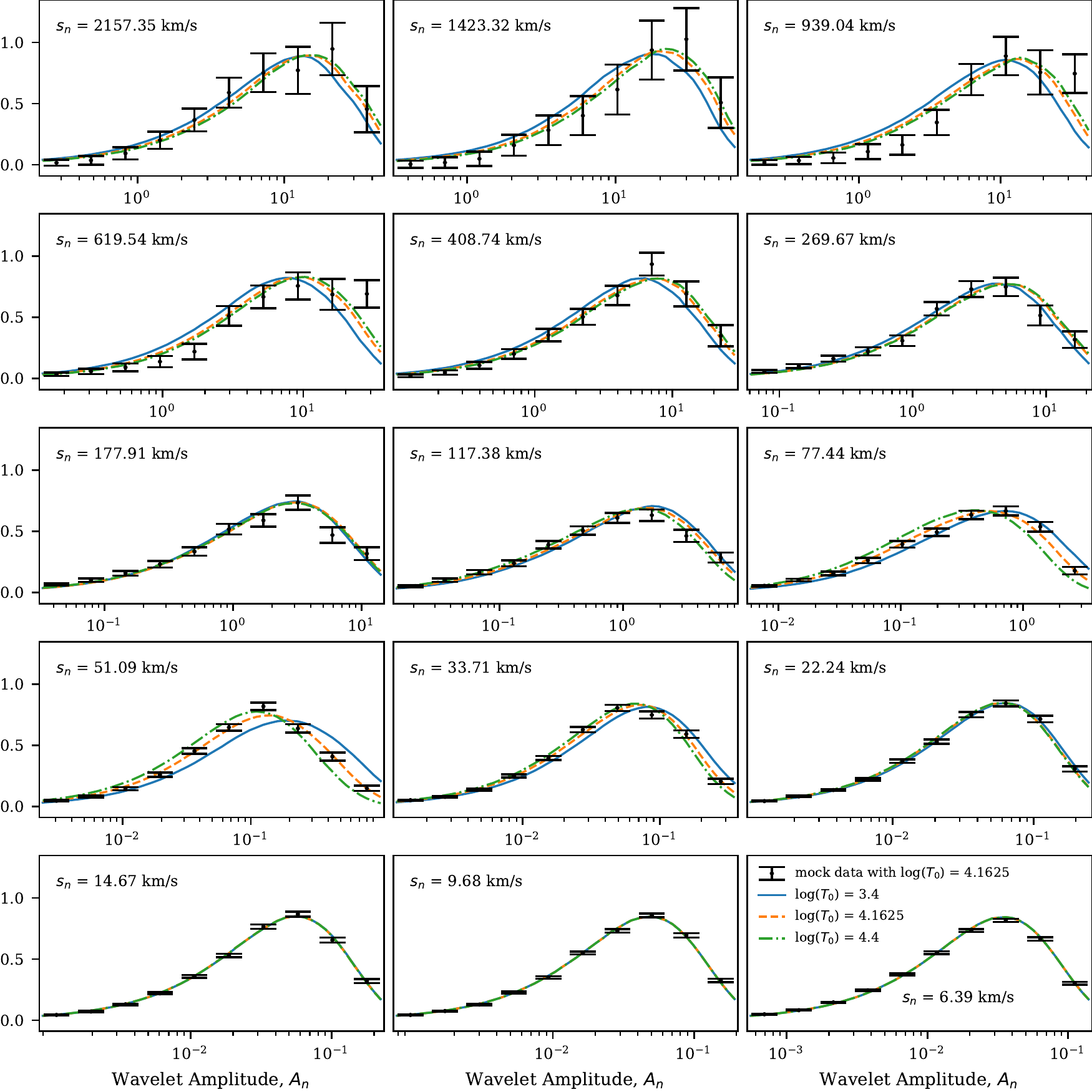}
    \caption{
        The black points show the PDFs from one mock data set for each $s_n$ with $\log(T_0) = 4.1265$.
        The $1\sigma$ error bars are calculated from the square root of the diagonal of the covariance matrix.
        Each panel also shows the "model" values of the PDFs from the stated smoothing scale for three different values of $T_0$: $\log(T_0) = 3.4$ (blue), $\log(T_0) = 4.1625$ (orange), and $\log(T_0) = 4.4$ (green).
        This figure qualitatively illustrates the ability of the wavelet PDF to differentiate between different $T_0$ values, which we formally quantify with Bayesian inference as discussed in Section \ref{section:stat inference}.
    }
    \label{fig:one_sn_mock_hist}
\end{figure*}

Figure \ref{fig:one_sn_mock_hist} shows the PDFs from one mock data set for each $s_n$ with $\log(T_0) = 4.1265$ (black points) and $z=5$.
An equivalent figure for $z=6$ can be found in Appendix \ref{fig:z6_power_mock}.
The $1\sigma$ error bars are calculated from the square root of the diagonal of the covariance matrix.
Each panel also shows the "model" values of the PDFs for three different values of $T_0$: $\log(T_0) = 3.4$ (blue), $\log(T_0) = 4.1625$ (orange), and $\log(T_0) = 4.4$ (green).
This figure qualitatively illustrates the ability of the wavelet PDF to differentiate between different $T_0$ values, which we formally quantify with Bayesian inference as discussed in Section \ref{section:stat inference}. 
The "model" PDFs for $s_n < \SI{22.24}{\kilo\meter\per\second}$ all overlap because the noise dominates the signal on these scales and all three PDFs are equivalent to the pure noise PDF. 

\begin{figure}
	\includegraphics[width=\columnwidth]{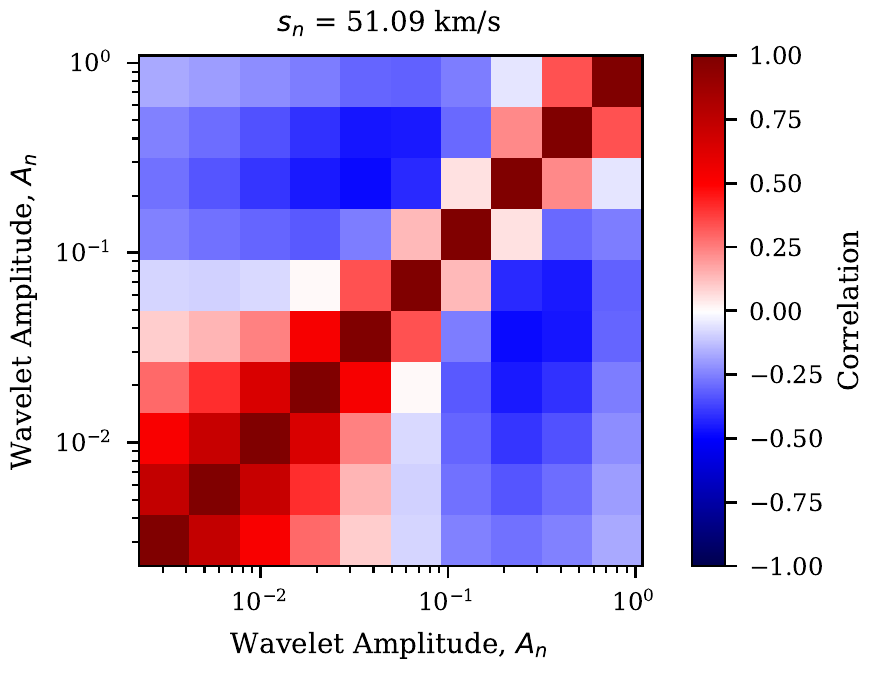}
    \caption{
        The correlation for the wavelet amplitude PDF for $\log(T_0) = 4.1625$ for $s_n = \SI{51.09}{\kilo\meter\per\second}$.
        There are positive correlations (red) between the bins at small wavelet amplitudes $A_n < 5 \times 10^{-2}$ with the other small values. 
        For larger values, there are negative correlations (blue) between the wavelet amplitudes $A_n > 0.1$ and all other wavelet amplitude values. 
        These effects are due to the shape of the PDF as well as the constraint that the PDF must integrate to 1. 
        }
    \label{fig:amp_correlation_one_sn}
\end{figure}

In order to understand the correlations present between the bins of a single wavelet PDF, we first calculate the model covariance matrix for $s_n = \SI{51.09}{\kilo\meter\per\second}$ and $\log(T_0) = 4.1265$ and then plot the correlation matrix in Figure \ref{fig:amp_correlation_one_sn}.
There are positive correlations (red) between the bins at small wavelet amplitudes $A_n < 5 \times 10^{-2}$ with the other small values. 
For larger values, there are negative correlations (blue) between the larger wavelet amplitudes $A_n > 0.1$ and all other wavelet amplitude values. 
These effects are due to the shape of the PDF as well as the constraint that the PDF must integrate to 1. 
Increasing the counts for any wavelet amplitude value will cause the counts in the peak of the PDF (around $A_n \sim 0.1$ as seen in Figure \ref{fig:one_sn_mock_hist} for $s_n = \SI{51.09}{\kilo\meter\per\second}$) to decrease due to the integral constraint on the PDF.
Meanwhile, the shape of the PDF means that when one bin along the tail ($A_n < 4 \times 10^{-2}$ as seen in Figure \ref{fig:one_sn_mock_hist} $s_n = \SI{51.09}{\kilo\meter\per\second}$) increases in counts, the other tail bins will increase as well. 

\begin{figure*}
	\includegraphics[width=\columnwidth*2]{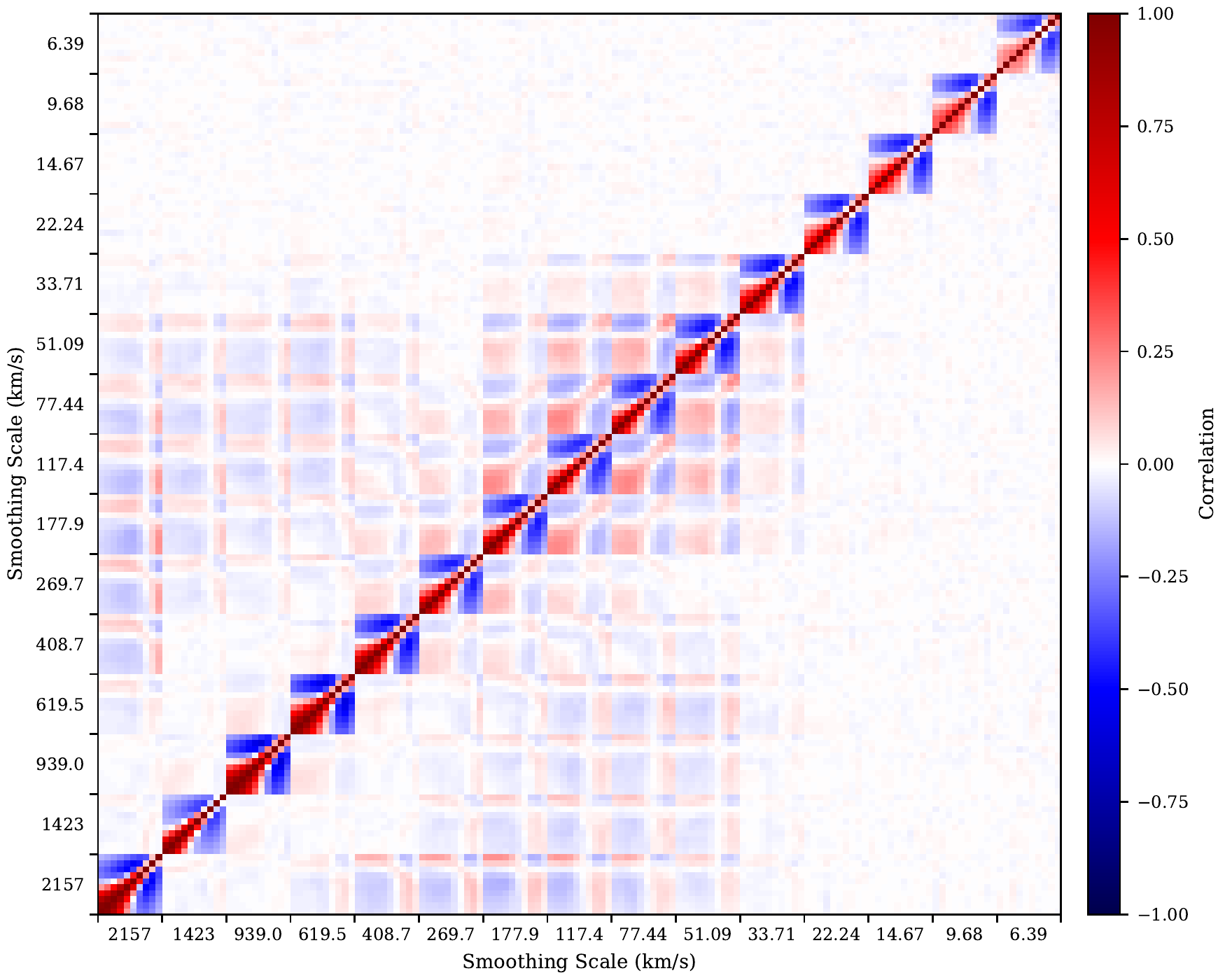}
    \caption{
        The correlation matrix for fifteen wavelet amplitude PDFs at $\log(T_0) = 4.1625$.
        The wavelet amplitude PDFs for large smoothing scales, $\SI{2157}{\kilo\meter\per\second} \geq s_n \geq \SI{33.7}{\kilo\meter\per\second}$, have significant correlations off the diagonal.
        The correlations between the PDFs with $\SI{177.9}{\kilo\meter\per\second} > s_n > \SI{33.71}{\kilo\meter\per\second}$ have the same pattern as the diagonal blocks modified by a small positive number (appearing mostly red). 
        The correlations between the PDFs for $s_n > \SI{408.7}{\kilo\meter\per\second}$ and $\SI{177.9}{\kilo\meter\per\second} > s_n > \SI{33.71}{\kilo\meter\per\second}$ have the same pattern as the diagonal blocks modified by a small negative number (appearing mostly blue). 
        For $s_n \leq \SI{22.2}{\kilo\meter\per\second}$, the wavelet amplitudes begin to be dominated by noise, so the correlations between the PDFs for different values of $s_n$ become very small. 
        This pattern mimics that seen in the power spectrum correlation shown in Figure \ref{fig:power_correlation}.
    }
    \label{fig:amp_correlation}
\end{figure*}

Ultimately, we will combine fifteen wavelet amplitude PDFs, each with a different value of $s_n$, in our measurement.
Our measurement will include the correlations between the different PDFs, unlike the measurements from both \citet{Lidz_2010} and \citet{gaikwad_2020} which ignore these correlations.
We include these correlations by using non-zero off diagonal terms in each covariance matrix, $\Sigma(T_0)$, when computing the likelihood in equation \eqref{eq:gauss_like}.
The correlations between different wavelet amplitude PDFs have never been considered in the previous literature on the Ly$\alpha$ forest. 
Our data vector is a concatenation of each wavelet amplitude PDF starting with the largest value of $s_n$ going down to the smallest value, making it $n_{\text{bins}} \times n_{s_n} = 10 \times 15 = 150$ points long. 
We expect that these correlations between different $s_n$ values will be non-negligible due to the spatial correlations shown in the periodogram (Figure \ref{fig:periodogram_clean}) as well as in the power spectrum (Figure \ref{fig:power_correlation}). 

In this case, the correlation matrix has dimension $150 \times 150$ and is shown in Figure \ref{fig:amp_correlation}. 
For visual purposes, the axes are labeled by the smoothing scale used to calculate the wavelet amplitude PDFs, but the correlations shown are between the wavelet amplitude bins (such as the labels in Figure \ref{fig:amp_correlation_one_sn}). 
Each $10 \times 10$ block along the diagonal is the correlation matrix for a single $s_n$ value. 
These diagonal blocks all appear very similar to the example shown for $s_n = \SI{51.09}{\kilo\meter\per\second}$ in Figure \ref{fig:amp_correlation_one_sn}, as expected from the similar shaped PDFs. 

The wavelet amplitude PDFs for large smoothing scales, $\SI{2157}{\kilo\meter\per\second} \geq s_n \geq \SI{33.7}{\kilo\meter\per\second}$, have significant correlations off the diagonal.
The correlations between the PDFs with $\SI{177.9}{\kilo\meter\per\second} > s_n > \SI{33.71}{\kilo\meter\per\second}$ have the same pattern as the diagonal blocks modified by a small positive number (appearing mostly red). 
The correlations between the PDFs for $s_n > \SI{408.7}{\kilo\meter\per\second}$ and $\SI{177.9}{\kilo\meter\per\second} > s_n > \SI{33.71}{\kilo\meter\per\second}$ have the same pattern as the diagonal blocks modified by a small negative number (appearing mostly blue). 
These modifications follow the same pattern as that in the correlation from the power spectrum shown in Figure \ref{fig:power_correlation} where there are positive correlations (red) between $4 \times 10^{-2} \SI{}{\second\per\kilo\meter} \lesssim k \lesssim \SI{0.2}{\second\per\kilo\meter}$ and negative correlations between $4 \times 10^{-2} \SI{}{\second\per\kilo\meter} \lesssim k \lesssim \SI{0.2}{\second\per\kilo\meter}$ and $k < 4 \times 10^{-2} \SI{}{\second\per\kilo\meter}$. 
This pattern arises from the underlying spatial correlations of the Ly$\alpha$ forest as was discussed for the power spectrum. 
For $s_n \leq \SI{22.2}{\kilo\meter\per\second}$, the wavelet amplitudes begin to be dominated by noise, so the correlations between the PDFs of different smoothing scales become very small.
This again mimics the behavior seen for $k > 0.2 \SI{}{\second\per\kilo\meter}$ in the power spectrum correlation matrix shown in Figure \ref{fig:power_correlation}.  

Many of these off diagonal covariance elements are very small, and it is challenging to measure small correlations from finite noisy data sets. 
For the full set of fifteen wavelet amplitude PDFs, the covariance matrix is 100 times larger than the power spectrum covariance matrix, making the calculation even more time consuming and difficult. 
This noise in the wavelet amplitude PDFs covariance matrix becomes quite noticeable in the posterior measurement on $T_0$. 
We reduce this noise by smoothing the covariance matrices over multiple thermal grid with one spline per matrix element. 
A detailed discussion of this smoothing can be found in Appendix \ref{section:appendix covariance smoothing}.

\subsection{Joint Wavelet-Power Likelihood} \label{section:combined likelihood}

\citet{gaikwad_2020} combined the wavelet amplitude PDFs with the power spectrum as well as the Doppler parameter distribution and curvature statistics to improve upon each of the individual measurements of the thermal state of the IGM. 
They did this by ignoring the correlations between PDFs for different smoothing scales as well as between the PDFs and the power spectrum despite the fact that these statistics were all measured from the same data set, and are thus surely correlated.
This application has motivated us to combine the power spectrum and wavelet amplitude PDFs while paying careful attention to correlations to see if this improves the precision of our mock measurement. 
We expect there to be non-negligible correlations between the wavelet amplitude PDFs and the power spectrum from equation \eqref{eq:power wavelet equivalence}, since this says the mean wavelet amplitude, i.e. the first moment of the wavelet PDF, contains the same information as a band power.

When combining the wavelet amplitude PDFs and the power spectrum, the data vector is the 150 element wavelet amplitude PDFs, i.e. 10 PDF bins $\times$ 15 smoothing scales discussed in Section \ref{section:amp likelihood}, with the addition of the 15 band-powers of $P_{\rm F}(k)$ added to the end. 
This makes the full data vector 165 points long and the correlation a complicated $165 \times 165$ matrix. 
To build intuition, we will first consider a subset of the full correlation matrix that consists of 
a single  wavelet amplitude PDF with the power spectrum. The correlation matrix in this situation will be only $25 \times 25$, i.e. 10 wavelet PDFs values and 15 band-powers.  

\begin{figure}
	\includegraphics[width=\columnwidth]{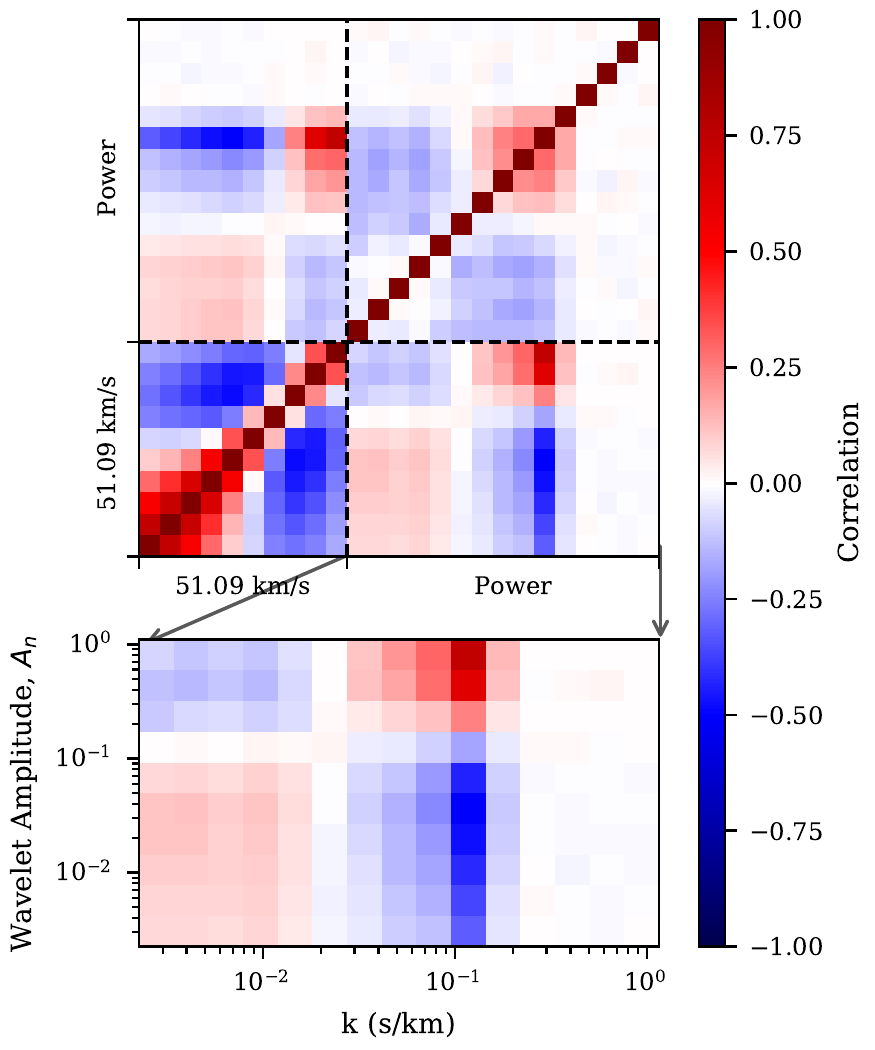}
    \caption{
        The top panel shows the correlation matrix for wavelet amplitude PDF with $s_n = \SI{51.09}{\kilo\meter\per\second}$ and the power spectrum for $\log(T_0) = 4.1625$.
        The dotted lines separate the part of the matrix that corresponds to the wavelet amplitude PDF (labeled by $s_n = \SI{51.09}{\kilo\meter\per\second}$) and the power spectrum (labeled "Power"). 
        The bottom left diagonal block is identical to the correlation matrix shown in Figure \ref{fig:amp_correlation_one_sn} while the top right diagonal is identical to the correlation matrix shown in Figure \ref{fig:power_correlation}. 
        The bottom right rectangle of the correlation matrix is blown up in the bottom panel, which shows the correlations between the power spectrum and the wavelet amplitude PDFs with the appropriate labels of $A_n$ and $k$. 
        The strongest correlations (both positive and negative) between the power and wavelet amplitude PDF are found in the column at $k = 6 / s_n = \SI{0.12}{\second\per\kilo\meter}$. 
        This $k$ value corresponds to the same scales probed by $s_n = \SI{51.09}{\kilo\meter\per\second}$.
        The behavior seen in the $k = \SI{0.12}{\second\per\kilo\meter}$ column of the bottom panel is replicated for the columns above $3 \times 10^{-2} \SI{}{\second\per\kilo\meter} \lesssim k \lesssim 0.2 \SI{}{\second\per\kilo\meter}$ modified by a small positive number.  
        The columns where $k < 4 \times 10^{-2} \SI{}{\second\per\kilo\meter}$ show the same behavior modified by a small negative number.
        These modifications mimic the pattern in the correlations for the power spectrum, in particular the positive (red) correlations for $3 \times 10^{-2} \SI{}{\second\per\kilo\meter} \lesssim k \lesssim 0.2 \SI{}{\second\per\kilo\meter}$ with $k = \SI{0.12}{\second\per\kilo\meter}$ and the negative (blue) correlations for $k < 4 \times 10^{-2} \SI{}{\second\per\kilo\meter}$ with $k = \SI{0.12}{\second\per\kilo\meter}$.
        }
    \label{fig:combined_correlation_s9}
\end{figure}

The correlation matrix for the wavelet amplitude PDF from $s_n = \SI{51.09}{\kilo\meter\per\second}$ and the power spectrum is shown in Figure \ref{fig:combined_correlation_s9}. 
The top panel has the full correlation matrix with the axes labeled by either the smoothing scale, $s_n = \SI{51.09}{\kilo\meter\per\second}$, which was used to calculate the $A_n$ or "Power" representing the different values of $k$.
The top right $15 \times 15$ diagonal block is identical to the correlation matrix for the power spectrum shown in Figure \ref{fig:power_correlation} and the bottom left $10 \times 10$ diagonal block is identical to the correlation matrix for one $s_n = \SI{51.09}{\kilo\meter\per\second}$ shown in Figure \ref{fig:amp_correlation_one_sn}. 

The off diagonal blocks show the correlations between the wavelet amplitude PDFs and the power spectrum. 
The bottom right rectangle of the correlation matrix is blown up in the bottom panel of the figure with the axes appropriately labeled by the wavelet amplitude $A_n$ from the PDF and the $k$ from the power spectrum.
The strongest correlations (both positive and negative) between the power and wavelet amplitude PDF are found in the column at $k = 6 / s_n = \SI{0.12}{\second\per\kilo\meter}$. 
This $k$ value corresponds to the same scales probed by $s_n = \SI{51.09}{\kilo\meter\per\second}$.
As the value of this $k$ bin increases, we expect the wavelet amplitude PDF to shift to higher values so that the average wavelet amplitude in the PDF increases, as required by equation \eqref{eq:power wavelet equivalence}. 
This shift causes the larger values of $A_n$ ($A_n > \SI{0.2}{\kilo\meter\per\second}$) to be
more common, resulting in a positive correlation (red) with larger PDF bins, while the smaller values of $A_n$ ($A_n < \SI{0.1}{\kilo\meter\per\second}$) are less common, resulting in a negative correlation (blue). 

The behavior seen in the $k = \SI{0.12}{\second\per\kilo\meter}$ column of the bottom panel is replicated for the columns above $3 \times 10^{-2} \SI{}{\second\per\kilo\meter} \lesssim k \lesssim 0.2 \SI{}{\second\per\kilo\meter}$ modified by a small positive number.  
The columns where $k < 4 \times 10^{-2} \SI{}{\second\per\kilo\meter}$ show the same behavior modified by a small negative number.
These modifications mimic the pattern in the correlations for the power spectrum, as shown in the upper right quadrant of the top panel and Figure \ref{fig:power_correlation}.
In particular, it replicates the positive (red) correlations for $3 \times 10^{-2} \SI{}{\second\per\kilo\meter} \lesssim k \lesssim 0.2 \SI{}{\second\per\kilo\meter}$ with $k = \SI{0.12}{\second\per\kilo\meter}$ and the negative (blue) correlations for $k < 4 \times 10^{-2} \SI{}{\second\per\kilo\meter}$ with $k = \SI{0.12}{\second\per\kilo\meter}$. 

\begin{figure*}
	\includegraphics[width=\columnwidth*2]{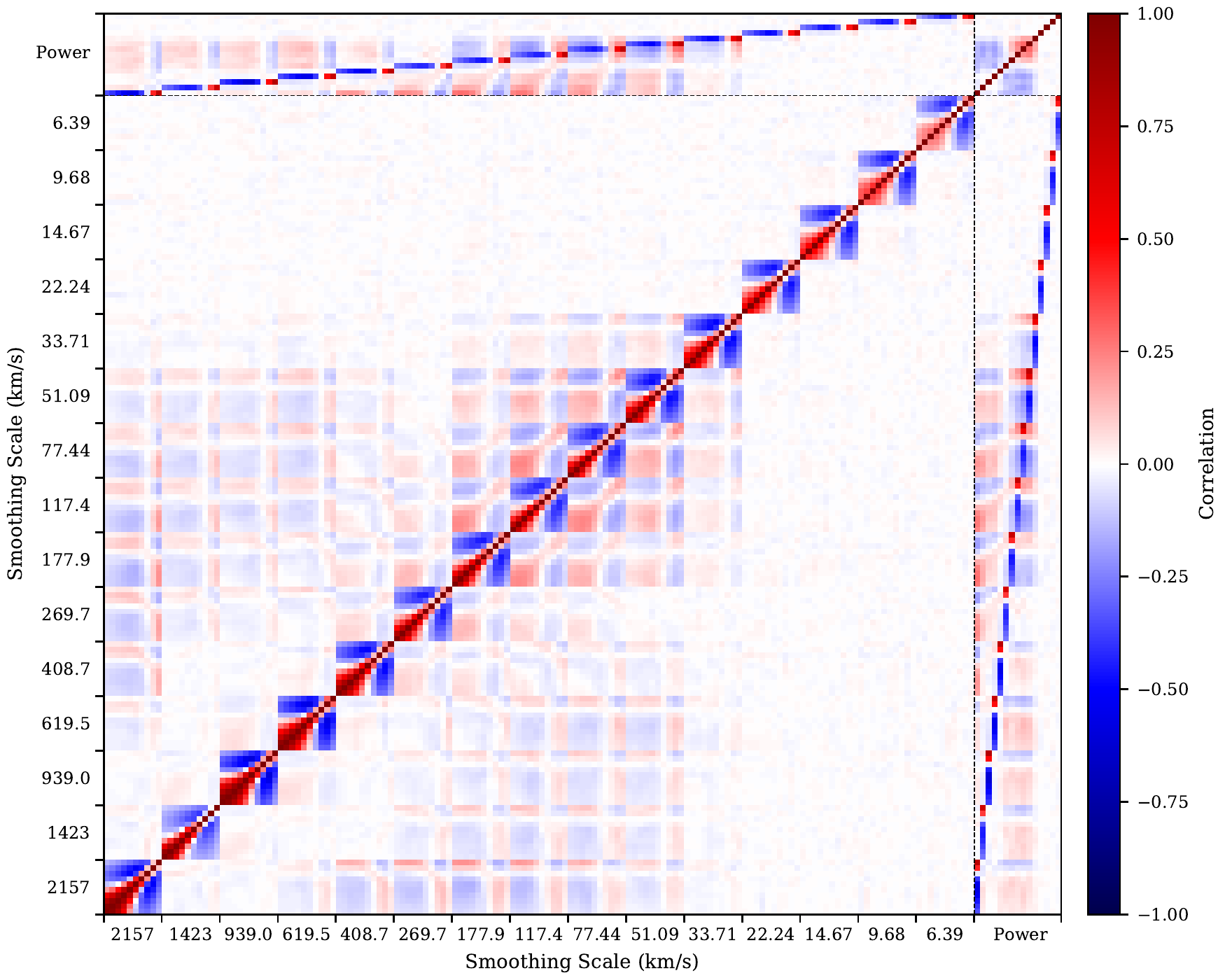}
    \caption{
    The correlation for all fifteen wavelet amplitude PDFs and the power spectrum combined for $\log(T_0) = 4.1625$.
    The axes are labeled by either the smoothing scale used to calculate $A_n$ or "Power" representing the $k$ bands. 
    The bottom left $150 \times 150$ diagonal block is identical to the correlation matrix for the wavelet amplitude PDFs shown in Figure \ref{fig:amp_correlation} while the top right $15 \times 15$ diagonal block is identical to the correlation matrix for the power spectrum shown in Figure \ref{fig:power_correlation}. 
    The right most column above "Power" shows similar behavior for each smoothing scale as was discussed for the bottom panel of Figure \ref{fig:combined_correlation_s9}. 
    The column with the strongest correlations for each smoothing scale always corresponds to $k = 6/s_n$ and the behavior in other columns above "power" follow the strongest bin modified by the correlations between the power bins. 
    }
    \label{fig:combined_correlation}
\end{figure*}

As discussed in the beginning of this section, the total data vector for the combination of the wavelet amplitude PDFs and the power spectrum is 165 points long. 
The full $165 \times 165$ correlation matrix is shown in Figure \ref{fig:combined_correlation}. 
The axes are labeled by either the smoothing scale used to calculate $A_n$ or "Power" representing the $k$ bands. 
The bottom left $150 \times 150$ diagonal block is identical to the correlation matrix for the wavelet amplitude PDFs shown in Figure \ref{fig:amp_correlation} while the top right $15 \times 15$ diagonal block is identical to the correlation matrix for the power spectrum shown in Figure \ref{fig:power_correlation}. 
The right most column above "Power" shows similar behavior for each smoothing scale as was discussed for the bottom panel of Figure \ref{fig:combined_correlation_s9}. 
The column with the strongest correlations for each smoothing scale always corresponds to $k = 6/s_n$ and the behavior in other columns above "power" follow the strongest bin modified by the correlations between the power bins. 

This data vector is larger than the data vector of the wavelet amplitude PDFs, which was discussed in Section \ref{section:amp likelihood}. 
Similarly, the noise in this covariance matrix is non-negligible and so we smooth the covariance matrix over the thermal grid with a spline in order to calculate the posteriors. 
This is discussed in more detail in Appendix \ref{section:appendix covariance smoothing}.


\section{Results} \label{section:results}

\subsection{$T_0$ Measurements} \label{sec:temp meas}

\begin{figure}
	\includegraphics[width=\columnwidth]{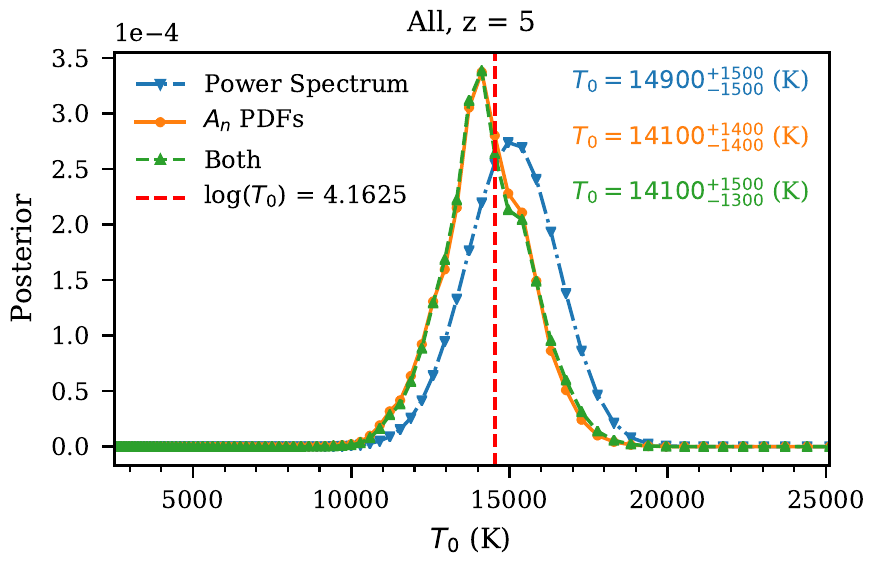}
    \caption{
        The posterior on $T_0$ for one mock data at $z=5$ set from three different methods: the power spectrum (blue triangles), the wavelet amplitude PDFs (orange circles), and the combination of the power spectrum and the wavelet amplitude PDFs (green triangles). 
        The mock data set used to calculate these posteriors is shown for the power in Figure \ref{fig:one_mock_power} and for the wavelet amplitude PDFs in Figure \ref{fig:one_sn_mock_hist}.
        The vertical dotted red line shows the true value of $T_0$ for the mock data set. 
        Qualitatively, the posterior from the power spectrum is less precise than the posterior from the wavelet amplitude PDFs and the combination of the power spectrum and wavelet amplitude PDFs do not improve the precision of the posterior over the wavelet amplitude PDFs alone. 
        The text in the corner is a quantitative measurement of $T_0$, giving the median value with the equivalent 1$\sigma$ errors for each method according to their colors in the same order as the legend.
    }
    \label{fig:posteriors}
\end{figure}

We can calculate the posterior probability of $T_0$ given a mock data set, $P(T_0 | \text{data})$, from equation \eqref{eq:bayes}.
In Figure \ref{fig:posteriors} we compare the posterior distribution of $T_0$ from one mock data set at $z=5$ from three different methods: the power spectrum (blue triangles), the wavelet amplitude PDFs (orange circles), and both the power spectrum and wavelet amplitude PDFs (green triangles). 
This mock data set is the same one shown in Figures \ref{fig:one_mock_power} and \ref{fig:one_sn_mock_hist}.
Visually, the wavelet amplitude PDFs provide a more precise posterior for $T_0$ than the power spectrum does while. 
The measurement of $T_0$ for these two methods are $T_0 = 14,900^{+1500}_{-1500}$ K (power spectrum) and $T_0 = 14,100^{+1400}_{-1400}$ K (wavelet amplitude PDFs). 
These errors are calculated by interpolating the cumulative distribution function (CDF) of the posterior onto the 15.9th and 84.1th percentiles, which correspond to the $1\sigma$ percentiles for a normal distribution. 
This region between these percentiles will be referred to as the $1\sigma$ region and the errors calculated from it as the equivalent $1\sigma$ errors throughout the end of this paper. 
The $1\sigma$ region is $7\%$ smaller for the wavelet amplitude PDFs posterior than the power spectrum posterior, showing that the wavelet amplitude PDF is more sensitive to $T_0$ than the power for this mock data set. 
Combining the power spectrum and the wavelet PDFs has a negligible effect on the posterior distribution resulting in $T_0 = 14,100^{+1500}_{-1300}$ K. 
This $1\sigma$ region has the same width as the one for the wavelet amplitude PDFs alone and so the combination does not improve the measurement's precision.

\begin{figure}
	\includegraphics[width=\columnwidth]{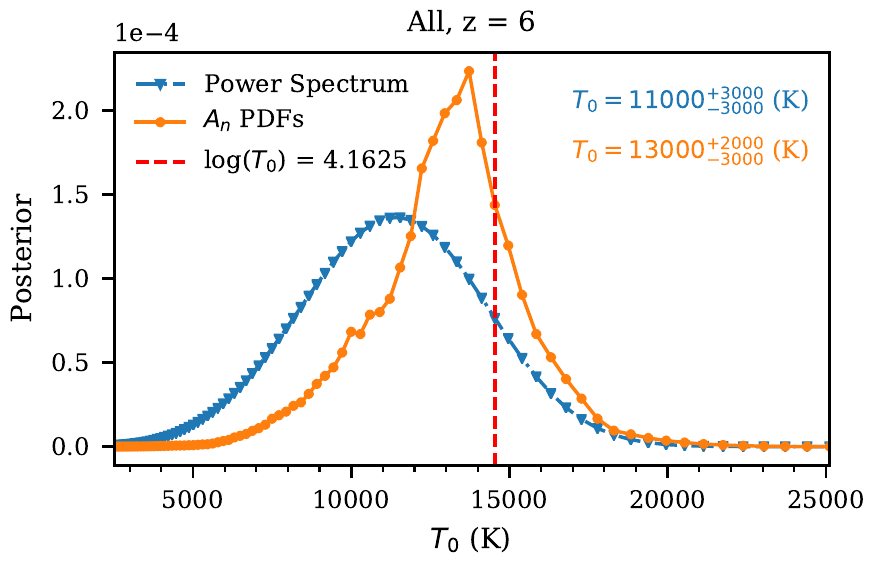}
    \caption{
        The posterior on $T_0$ for one mock data set at $z=6$ from two different methods: the power spectrum (blue triangles) and the wavelet amplitude PDFs (orange circles).
        The mock data set used to calculate these posteriors is shown for the power in Figure \ref{fig:z6_power_mock} and for the wavelet amplitude PDFs in Figure \ref{fig:z6_amps_mock}.
        The vertical dotted red line shows the true value of $T_0$ for the mock data set. 
        Qualitatively, the posterior from the power spectrum is less precise than the posterior from the wavelet amplitude PDFs. 
        The text in the corner is a quantitative measurement of $T_0$, giving the median value with the equivalent 1$\sigma$ errors for each method according to their colors in the same order as the legend.
    }
    \label{fig:z6_posterior}
\end{figure}

In Figure \ref{fig:z6_posterior} we compare the posterior distribution of $T_0$ from one mock data set at $z=6$ from two different methods: the power spectrum (blue triangles) and the wavelet amplitude PDFs (orange circles). 
This mock data set is the same one shown in Figures \ref{fig:z6_power_mock} and \ref{fig:z6_amps_mock}.
The measurement of $T_0$ for these two methods are $T_0 = 11,000^{+3000}_{-3000}$ K (power spectrum) and $T_0 = 13,000^{+2000}_{-3000}$ K (wavelet amplitude PDFs) where the errors are calculated in the same way as they were for $z=5$. 
At this redshift, the wavelet amplitude PDF measurements resulted in a 20\% improvement of the $1\sigma$ errors when compared to the results from power spectrum measurement from the same mock data, almost three times the improvement seen at $z=5$, though here the errors only have one significant figure so improvement is a coarser measurement.

To further quantify the difference in the precision of these posteriors, we calculated the equivalent $1\sigma$ errors for $1,000$ mock data sets at $\log(T_0) = 4.1265$ for both $z=5$ and $z=6$. 
These resulting mean and variance of these values are listed in Table \ref{tab:results}. 
On average, the posteriors for the wavelet amplitude PDFs at $z=5$ are $7\%$ smaller than those from the power spectrum, again showing that the wavelet amplitude PDFs are more sensitive on average than the power spectrum.  
However, the variance on these power spectrum errors are $75\%$ smaller, meaning the power spectrum posteriors are more consistently large while the wavelet amplitude PDFs vary more in size. 
The average errors on the posteriors from combining both the wavelet amplitude PDFs and the power spectrum show no improvement over the errors from the wavelet amplitude PDFs alone again showing this combination does not improve the measurement. 

For $z=6$, the posteriors for the wavelet amplitude PDFs are $12\%$ smaller than those from the power spectrum while the variance on the power spectrum errors are $67\%$ smaller.
This means that the wavelet amplitude PDFs are again more sensitive on average but the errors vary more than the power spectrum. 
This agrees with the results at $z=5$, though again we find that the wavelets lead to an even greater improvement on the average sensitivity by a factor of two. 
Physically, at this higher redshift, more of the spectra consists of absorption troughs, giving the wavelet amplitude PDFs even greater potential to improve on the power spectrum measurements since they maintain spatial information.

\begin{table}
    \centering
    \caption{
        This table shows the mean and variance of the 1$\sigma$ values from 1,000 mock data sets at $\log(T_0) = 4.1265$ at both $z=5$ and $z=6$. 
        The $1\sigma$ errors for the wavelet amplitudes PDFs are on average $7\%$ smaller at $z=5$ and $12\%$ smaller at $z=6$ than those for the power spectrum, though they do have a higher variance. 
        The error calculated from combining the power spectrum and the wavelet analysis PDFs at $z=5$ does not improve the errors on average over the wavelet amplitude PDFs alone.
        }
	\label{tab:results}
    \begin{tabular}{|l|l|l|l|}
        \hline
        $z$ & Method                 & $\overline{\sigma}_{+}$ & $\overline{\sigma}_{-}$ \\ \hline
        \multirow{3}{*}{5} & Power Spectrum         & $1490 \pm 50$           & $-1520 \pm 50$          \\
                           & Wavelet Amplitude PDFs & $1400 \pm 200$          & $-1400 \pm 200$         \\
                           & Both                   & $1400 \pm 200$          & $-1400 \pm 200$         \\ \hline
        \multirow{2}{*}{6} & Power Spectrum         & $3030 \pm 190$          & $-3140 \pm 200$         \\
                           & Wavelet Amplitude PDFs & $2700 \pm 600$          & $-2700 \pm 600$         \\ \hline
    \end{tabular}
\end{table}

\subsection{Inference Testing} \label{section:inference testing}

In order to test the fidelity of our statistical inference procedure and results, we perform an inference test. 
The goal of this test is to check that this calculated posterior behaves as a posterior probability should: if the true value of $T_0$ for the mock data falls into the equivalent of the 1$\sigma$ region of the posterior $\sim 68\%$ of the time (and the 2$\sigma$ region 95\% of the time). 
We again calculate these equivalent 1$\sigma$ and 2$\sigma$ regions for our posteriors in the same way as discussed in Section \ref{sec:temp meas}. 
We integrate the posterior to get the CDF onto the 15.9th and 84.1th percentiles for $1\sigma$ and onto the 2.3rd and 97.7th percentile for $2\sigma$.
These percentiles correspond to the $1\sigma$ and $2\sigma$ percentiles for a normal distribution. 
From here, we count the number of times the true value of $T_0$ fell into these regions region. 
Ideally, the true value of $T_0$ should fall into the $1\sigma$ region $68.3\%$ of the time and it should fall into the $2\sigma$ region $95.4\%$ of the time.

We did this for 1,000 mock data sets at three different values of $T_0$ for $z=5$ and one value of $T_0$ at $z=6$.
We chose to only look at one value of $T_0$ at the higher redshift because the posteriors are broader and we are more likely to run into edge effects at the other $T_0$ values.
The errors are calculated by $\sqrt{N}/10000$ where $N$ is the number of times the true value fell into the desired region and 1,000 is the total number of mocks used. 
The results, shown in Table \ref{tab:inference}, are consistent with the expected values of $68.3\%$ and $95.4\%$ within the calculated errors and we pass this inference test. 

\begin{table}
    \centering
    \caption{
        This table shows the results of our inference test for three values of $T_0$ and three statistical methods (power spectrum, wavelet amplitude PDFs, and the combination of the two) at $z=5$. 
        We have also included our results for one value of $T_0$ and two statistical methods (power spectrum and wavelet amplitude PDFs at $z=6$.
        We calculated the equivalent 1$\sigma$ and 2$\sigma$ regions from the CDF and then determined the frequency with which the true $T_0$ values fell into these regions. 
        These results are presented for 1,000 mock data sets and are consistent with a true distribution function with our expected errors.
        }
	\label{tab:inference}
    \begin{tabular}{|l|l|l|l|l|}
        \hline
        $z$                & Method                                  & $\log(T_0)$ & \% in 1$\sigma$    & \% in 2$\sigma$ \\ \hline
        \multirow{9}{*}{5} & \multirow{3}{*}{Power Spectrum}         & 3.9         & $70.0 \pm 2.6$     & $94.6 \pm 3.1$  \\ \cline{3-5} 
                           &                                         & 4.1265      & $68.9 \pm 2.6$     & $96.0 \pm 3.1$  \\ \cline{3-5} 
                           &                                         & 4.2875      & $68.7 \pm 2.6$     & $96.2 \pm 3.1$  \\ \cline{2-5}
                           & \multirow{3}{*}{Wavelet Amplitude PDFs} & 3.9         & $63.5 \pm 2.5$     & $94.0 \pm 3.1$  \\ \cline{3-5} 
                           &                                         & 4.1265      & $67.6 \pm 2.6$     & $95.0 \pm 3.1$  \\ \cline{3-5} 
                           &                                         & 4.2875      & $69.3 \pm 2.6$     & $95.9 \pm 3.1$  \\ \cline{2-5}
                           & \multirow{3}{*}{Both}                   & 3.9         & $63.7 \pm 2.5$     & $93.1 \pm 3.1$  \\ \cline{3-5} 
                           &                                         & 4.1265      & $67.1 \pm 2.6$     & $94.1 \pm 3.1$  \\ \cline{3-5} 
                           &                                         & 4.2875      & $68.5 \pm 2.6$     & $95.3 \pm 3.1$  \\ \hline
        \multirow{2}{*}{6} & Power Spectrum                          & 4.1265      & $68.1 \pm 2.6$     & $95.6 \pm 3.1$  \\ \cline{2-5}
                           & Wavelet Amplitude PDFs                  & 4.1265      & $65.7 \pm 2.6$     & $93.6 \pm 3.1$  \\ \hline
    \end{tabular}
\end{table}

\subsection{Ignoring Correlations} \label{sec:ignoring_correlations}

\begin{figure}
	\includegraphics[width=\columnwidth]{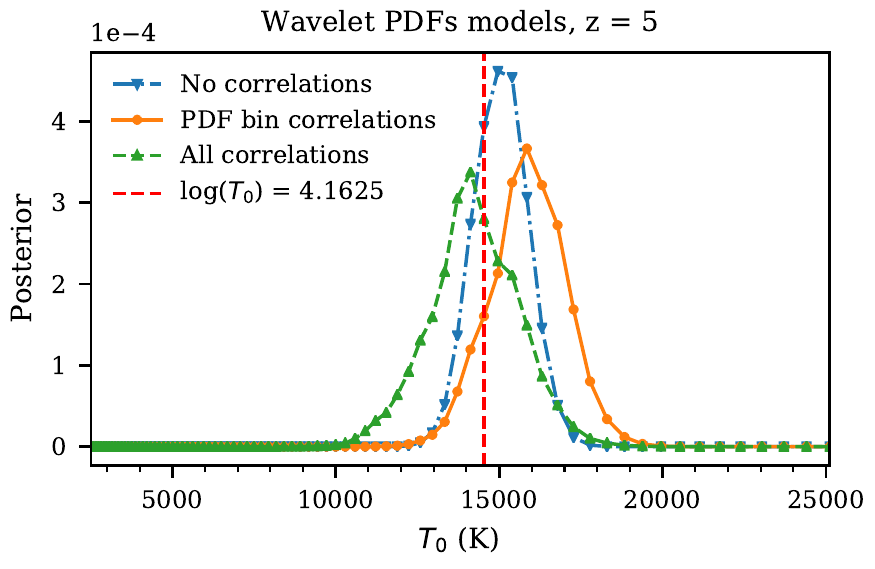}
    \caption{
        The posterior on $T_0$ for one mock data set for the wavelet amplitude PDFs using three different covariance matrices. 
        The three matrices are described in more details in Section \ref{sec:ignoring_correlations}.
        They are: a diagonal-only covariance matrix which includes no correlations (blue triangles), a diagonal-block matrix that only contains correlations between PDF bins for the same wavelet scale (orange circles), and the full covariance matrix with all correlations (green triangles). 
        The posterior from the diagonal matrix which has no correlations is much more narrow than the other two posteriors which are roughly the same width and height as each other. 
        }
    \label{fig:posteriors_amps_ignore}
\end{figure}

We further investigated the posterior distributions from the wavelet amplitude PDFs and the combined wavelet amplitude PDFs and power spectrum measurements at $z=5$ while ignoring certain correlations. 
To begin, we considered the posterior from the wavelet amplitude PDFs alone. 
We constructed three distinct covariance matrices from our initial full calculation, which is shown in Figure \ref{fig:amp_correlation}. 
The first covariance matrix considered is made up of the same values along the diagonal and zeros for all off-diagonal elements. 
This is similar to the covariance considered in \citet{Lidz_2010} and is referred to as the ``no correlations'' model in Figure \ref{fig:posteriors_amps_ignore} and Table \ref{tab:inference_ignore}. 
Next we construct a covariance matrix that includes the correlations between bins of the individual wavelet amplitude PDFs but ignores the correlations between different values of $s_n$.
The resulting correlation matrix would have the same values for the fifteen $10 \times 10$ diagonal blocks in Figure \ref{fig:amp_correlation} and zeros at all other locations. 
This is the similar to the covariance model considered in \citet{gaikwad_2020} and is referred to as the ``PDF bin correlations" model in Figure \ref{fig:posteriors_amps_ignore} and Table \ref{tab:inference_ignore}.
Finally we considered the full covariance matrix presented in this work in Figure \ref{fig:amp_correlation}, which we have labeled as ``all correlations" in Figure \ref{fig:posteriors_amps_ignore} and Table \ref{tab:inference_ignore}.
The resulting posteriors from these three models is shown in Figure \ref{fig:posteriors_amps_ignore}.
For the mock dataset shown in this figure (which is the same one shown throughout the rest of the paper) the ``no correlations" model reduces the width of the posterior while the ``PDF bin correlations" remains a similar width when compared to ``all correlations".
The median value does not agree for any of these posteriors though the whole distribution of the posteriors have significant overlap.
We also performed the same inference test on the posteriors for these models which will be discussed at the end of this section with results in Table \ref{tab:inference_ignore}. 

\begin{figure}
	\includegraphics[width=\columnwidth]{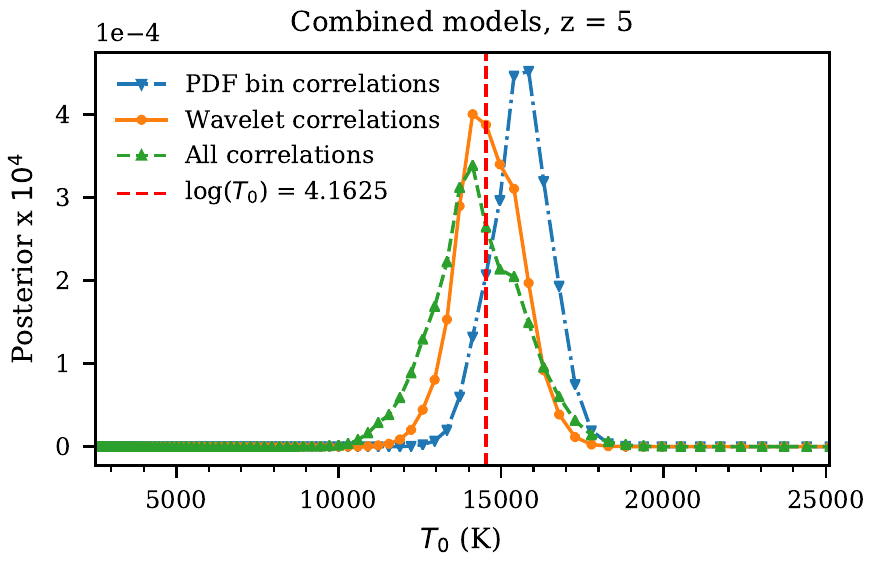}
    \caption{
        The posterior on $T_0$ for one mock data set for the combination of wavelet amplitude PDFs and the power spectrum using three different covariance matrices. 
        The three matrices are described in more details in Section \ref{sec:ignoring_correlations}.
        They are: a diagonal-block matrix with 16 distinct blocks for the wavelet PDF bin correlations and power correlations separately (blue triangles), a diagonal-block matrix with 2 distinct diagonal blocks that contains the full wavelet correlations and the power correlations separately (orange circles), and the full covariance matrix with all correlations including the cross-correlations between the wavelet and the power spectrum (green triangles). 
        Adding additional correlations caused the posterior distribution to broaden each time. 
        }
    \label{fig:posteriors_combined_ignore}
\end{figure}

Next, we again constructed three different covariance matrices for the combination of the wavelet amplitude PDFs and the power spectrum where the initial covariance matrix is shown in Figure \ref{fig:combined_correlation}.
First we considered only the correlations between PDF bins for the wavelet amplitude and the full correlations for the power spectrum.
The resulting covariance matrix has fifteen $10 \times 10$ diagonal blocks followed by one $15 \times 15$ diagonal block and zeros at all other locations.
This is similar to how the combination of different wavelet scales and different statistics were done in \citet{gaikwad_2020}.
We refer to this model as the ``PDF bin correlations" model in Figure \ref{fig:posteriors_combined_ignore} and Table \ref{tab:inference_ignore}.
The subset of this matrix for the wavelet PDFs matches that used in Figure \ref{fig:posteriors_amps_ignore} with the same name.
Next we consider the full wavelet correlations for the PDF bins as well as the different wavelet scales combined with the full power correlations but ignoring all cross correlations. 
The resulting matrix would have two diagonal blocks: one for the wavelet amplitude PDFs that has dimensions $150 \times 150$ and is shown in Figure \ref{fig:amp_correlation}, and one for the power spectrum that has dimensions $15 \times 15$ and is shown in Figure \ref{fig:power_correlation}.
We refer to this model as the ``wavelet correlations" model in Figure \ref{fig:posteriors_combined_ignore} and Table \ref{tab:inference_ignore}.
We also again considered the full covariance matrix presented in this work in Figure \ref{fig:combined_correlation}, which we have labeled as ``All correlations" in Figure \ref{fig:posteriors_combined_ignore} and Table \ref{tab:inference_ignore}.
The resulting posteriors from these three models is shown in Figure \ref{fig:posteriors_combined_ignore}.
For the mock dataset shown in this figure (which is the same one shown throughout the rest of the paper) the ``PDF bin correlations" has the most narrow posterior while the ``wavelet correlations" is more narrow than ``all correlations" but broader than ``PDF bin correlations".
Again, each posterior has a shifted median value compared to the others, though all the posteriors have significant overlap with each other.

We repeat the inference test described in section  \ref{section:inference testing} for these models and have presented the results in Table \ref{tab:inference_ignore}.
For the wavelet amplitude PDF models, only the ``all correltions'' model passed our inference test with the true value of $T_0$ falling in our 1$\sigma$ region for 67.6\% of mock posteriors and the true value of $T_0$ falling in our 2$\sigma$ region for 95.0\% of mock posteriors.
In comparison, the ``no correlations" model only had the true value of $T_0$ falling in our 1$\sigma$ region for 35.8\% of mock posteriors and the true value of $T_0$ falling in our 2$\sigma$ region for 62.7\% of mock posteriors.
We can roughly estimate that, since $68 / 36 \sim 1.9$, we would need to widen the posterior for the ``no correlations" model by a factor of 1.9 to pass the inference test. 
If we take the ``all correlations" model as the true information contained the in wavelet amplitude PDFs then the ``no correlations" posterior needs to be both shifted and broadened to a lesser extent to match this posterior. 
A similar calculation can be done for the ``PDF bin correlations" model which is more similar to the ``all correlations" model initially. 

The results for the inference test on the different models of the combined wavelet amplitude PDF and power spectrum correlations are also shown in Table \ref{tab:inference_ignore}.
Here, again, only the ``all correlations" model passed our inference test with the true value of $T_0$ falling in our 1$\sigma$ region for 67.1\% of mock posteriors and the true value of $T_0$ falling in our 2$\sigma$ region for 94.1\% of mock posteriors.
In comparison, the ``PDF bin correlations" model only had the true value of $T_0$ falling in our 1$\sigma$ region for 47.7\% of mock posteriors and the true value of $T_0$ falling in our 2$\sigma$ region for 79.4\% of mock posteriors.
We can roughly estimate that, since $68 / 48 \sim 1.4$, we would need to widen the posterior for the ``PDF bin correlations" model by a factor of 1.4 to pass the inference test. 
If we take the ``all correlations" model as the true information contained the in combination of the wavelet amplitude PDFs and the power spectrum then the ``PDF bin correlations" posterior needs to be both shifted and broadened to a lesser extent to match this posterior. 
A similar calculation can be done for the ``wavelet correlations" model which is more similar to the ``all correlations" model to start. 

\begin{table*}
    \caption{
        This table shows the results of our inference test when ignoring correlations for either the wavelet amplitude PDFs or the combination of the wavelet amplitude PDFs and the power spectrum. 
        The three models of correlations for each statistic is described in Section \ref{sec:ignoring_correlations}. 
        The inference test was done for only one true value of $T_0$, $\log(T_0) = 4.1265$.
        We calculated the equivalent 1$\sigma$ and 2$\sigma$ regions from the CDF and then determined the frequency with which the true $T_0$ values fell into these regions for 1,000 mock data sets. 
        Only the models that considered all the correlations as presented in this paper labeled as ``all correlations" for each statistic passed the inference test.
        The other two models for both statistics did not recover the true value of $T_0$ the expected number of times. 
        }
    \label{tab:inference_ignore}
    \begin{tabular}{|l|l|l|l|}
        \hline
        Method                                  & Correlations              & \% in 1$\sigma$ & \% in 2$\sigma$ \\ \hline
        \multirow{3}{*}{Wavelet Amplitude PDFs} & No correlations           & $35.8 \pm 1.9$  & $62.7 \pm 2.5$  \\ \cline{2-4} 
                                                & PDF bin correlations      & $55.2 \pm 2.3$  & $87.0 \pm 2.9$  \\ \cline{2-4} 
                                                & All Correlations          & $67.6 \pm 2.6$  & $95.0 \pm 3.1$  \\ \hline
        \multirow{3}{*}{Both PDFs and Power}    & PDF bin correlations      & $47.7 \pm 2.2$  & $79.4 \pm 2.8$  \\ \cline{2-4} 
                                                & Wavelet correlations      & $55.3 \pm 2.4$  & $86.5 \pm 2.9$  \\ \cline{2-4} 
                                                & All Correlations          & $67.1 \pm 2.6$  & $94.1 \pm 3.1$  \\ \hline
    \end{tabular}
\end{table*}

\subsection{Comparison to Previous Work}

\citet{Lidz_2010} made a measurement of the thermal state of the IGM with one wavelet amplitude PDF while ignoring the correlations between bins of the PDF. 
This measurement of $T_0$ is higher than those from most other statistics and also has larger error bars. 
This is shown in \citet{walther_2019}, which used the same quasar data set as \citet{Lidz_2010} when measuring the power spectrum but added additional data to roughly double the data set size. 
Figure 15 of \citet{walther_2019} shows the resulting thermal state constraints compared to \citet{Lidz_2010} as well as other measurements. 
We investigated the effect of ignoring the correlations between PDF bins in Section \ref{sec:ignoring_correlations} and found that this would result in underestimated errors. 
This would imply that the error bars from \citet{Lidz_2010} would not reflect the true precision of the measurement. 
Additionally, Figure \ref{fig:posteriors_amps_ignore} shows that ignoring these correlations also shifts the peak of the posterior. 
However, we did not consider only using one smoothing scale as was done in \citet{Lidz_2010}, which we would expect to broaden the posterior of the measurement. 
Thus we can not precisely estimate how shifted or underestimated the measurement and errors from \citet{Lidz_2010} would have been. 

\citet{gaikwad_2020} made a measurement of the thermal state of the IGM using eight wavelet amplitude PDFs with $\SI{30}{\kilo\meter\per\second} \leq s_n \leq \SI{100}{\kilo\meter\per\second}$, the power spectrum, and by combining these statistics along with the Doppler parameter distribution and curvature statistics.
All of these statistics were calculated from the same data set and the correlations between PDF bins of different smoothing scale and the individual statistics were ignored. 
The main results are shown in Figure 14 of \citet{gaikwad_2020}, where the different statistics have measurements that agree with each other but differ in error bar size and the joint constraints visually appear to have the smallest error bars most frequently. 
Our analysis has shown that combining the wavelet amplitude PDFs and the power spectrum from the same data set does not improve the measurement of $T_0$ on average.
However, we did not further investigate the additional statistics that \citet{gaikwad_2020} considered nor did we consider the situation where the smoothing scales used to calculate the wavelet amplitudes did not span the range of $k$ values considered in the power measurement. 
For these reasons, our work here varies considerably from the work done by \citet{gaikwad_2020}.
It is likely that combining the power spectrum and wavelet amplitude PDFs when there isn't a full correspondence between $s_n$ and $k$ would lead to an improvement of the measurement from either statistic alone.
However, as long as some $k$ and $s_n$ values overlap we expect there to be non-negligible correlations that were been ignored in \citet{gaikwad_2020}. 
The two additional statistics are also likely to improve the combined measurement beyond the combination of only power spectrum and wavelet amplitude PDFs.
Adding these statistics to the work presented here would require calculating the correlations between the different statistics as we did between the power spectrum and wavelet amplitude PDFs. 
Calculating these statistics and exploring the relevant correlations is beyond the scope of this work. 
For these reasons we can not precisely estimate the correct size of the error bars from their combined measurement.

We did consider the effect of ignoring the correlations between PDFS from different smoothing scales on the posterior for the wavelet amplitude PDFs alone in Section \ref{sec:ignoring_correlations} and Figure \ref{fig:posteriors_amps_ignore}. 
We found that ignoring these correlations caused the posterior to shift and underestimate the errors (the orange and green lines in Figure \ref{fig:posteriors_amps_ignore}). 
Using our inference test in Table \ref{tab:inference_ignore}, for the wavelet amplitude PDFs with ``PDF bin correlations" only the true value of $T_0$ fell in the 1$\sigma$ region $55\%$.
This would imply the need to grow the 1$\sigma$ region by a factor of $68/55 \sim 1.24$ or a $24\%$ increase.
It is therefore likely that the errors on the measurement from only the wavelet amplitudes in \citet{gaikwad_2020} are underestimated. 

Overall, Figures \ref{fig:amp_correlation_one_sn}, \ref{fig:amp_correlation}, \ref{fig:combined_correlation_s9}, and \ref{fig:combined_correlation} show that the off-diagonal terms in the covariance matrix are non-negligible and should be included in future analysis using wavelet amplitudes in order to achieve accurate error estimates.  
Figures \ref{fig:posteriors_amps_ignore} and \ref{fig:posteriors_combined_ignore} and Section \ref{sec:ignoring_correlations} additionally demonstrate and discuss the effects of ignoring these correlations on the posteriors.


\section{Conclusion} \label{section:conclusion}

We have expanded upon the wavelet analysis methods used by \citet{Lidz_2010} and \citet{gaikwad_2020} to study the thermal state of the IGM. 
Our method combines fifteen wavelet amplitude PDFs with smoothing scales that span the full range of scales probed by $P_{\rm F}(k)$ and, for the first time, provides a full accounting of the correlations between these PDFs. 
We also calculated $P_{\rm F}(k)$ from the same simulated data in order to compare the precision of measurements on $T_0$ from these statistics. 
In order to rigorously combine the wavelet amplitude PDFs and power spectrum, we calculated the cross-correlations between $P_{\rm F}(k)$ and the wavelet amplitude PDFs. 
We presented examples of each of these correlation matrices in Figures \ref{fig:amp_correlation_one_sn}, \ref{fig:amp_correlation}, \ref{fig:combined_correlation_s9}, and \ref{fig:combined_correlation}.
Figures \ref{fig:amp_correlation} and \ref{fig:combined_correlation} showed the non-negligible off-diagonal correlations between the different smoothing scales and the different statistics. 
With our method at $z=5$, the posterior of $T_0$ using the wavelet amplitude PDFs is on average $7\%$ more precise than the power spectrum measurement on the same data. 
This means getting the same precision measurement with the power spectrum requires $\sim 15\%$ more data.
Combining the power spectrum and wavelet amplitude PDFs did not significantly improve in precision of the posterior on $T_0$ over that from the wavelet amplitude PDFs alone, indicating that they contain the same information. 
At $z=6$ we found that the posterior of $T_0$ using the wavelet amplitude PDFs is on average $12\%$ more precise which would require $\sim 15\%$ more data to achieve the same accuracy with the power spectrum.
Additionally, we calculated posteriors on $T_0$ at $z=5$ with covariance matrices that ignored the off-diagonal correlations between PDF bins, between smoothing scales, and between the different statistics in Figures \ref{fig:posteriors_amps_ignore} and \ref{fig:posteriors_combined_ignore}. 
We were unable to pass an inference test with these posteriors (as reported in Table \ref{tab:inference_ignore}) which implies that the errors are underestimated in these cases.
This further demonstrated the significance of the off-diagonal terms in the covariance matrices and that they must be computed for a robust statistical analysis.

Here we adopted a simple model of the thermal state of the IGM which depended on a single parameter $T_0$.
For the more common and general case of multiple model parameters, the wavelet amplitude PDFs have even greater potential to better constrain model parameters when compared to the 1D flux power spectrum. 
To reiterate, the wavelet amplitude PDF characterizes the full wavelet distribution while the 1D flux power spectrum contains information on the mean of the wavelet amplitude PDF, see Equation \eqref{eq:power wavelet equivalence}. 
If we are only varying one model parameter and this parameter shifts the wavelet amplitude PDFs, the mean of the wavelet PDF may effectively contain all the information on the differences of the model.
This is true in our thermal model with $T_0$, as can be seen in Figures \ref{fig:two_panel_amp_pdf} and \ref{fig:one_sn_mock_hist}.
In more sophisticated models, like those of reionization \citep{Onorbe_2019, Boera_2019} and WDM \citep{Viel_2013, Irsic_2017}, we would want to vary multiple model parameters (such as $\gamma$, $\langle F \rangle$, and $m_{\text{WDM}}$).
Multiple model parameters are likely to cause changes in the full distribution of wavelet amplitudes beyond shifts in the mean. 
Wavelet amplitude PDF are sensitive to these additional changes while the power spectrum is not, meaning the wavelet amplitude PDFs could better discriminate between models to an even greater extent than they do for only one model parameter when compared to the power spectrum.
Investigation of models with multiple thermal parameters is beyond the scope of this paper but is a more realistic and promising area to explore wavelet analysis. 

Wavelets are an independent statistic that can be used to probe the small-scale structure of the IGM through the Ly$\alpha$ forest. 
They can be used as a check on alternative statistics such as the power spectrum, Doppler parameter distribution, and curvature statistics since, in principle, each statistic may be sensitive to different systematics. 
In addition, the wavelet amplitude PDFs are higher precision than the power spectrum. 
Wavelets have an added benefit of providing Fourier information in configuration space which may be useful in other areas, such as in quasar proximity zones \citep{Khrykin_2016} or when looking for temperature fluctuations in the IGM
\citep{theuns_zaroubi_2000, theuns_2002_fluct, Zaldarriaga_2002, fang_2004, Lai_2006, McQuinn_2011}. 
We have not studied the effects of a late ending reionization with remaining temperature fluctuations or a varying UVB background on the shape of the wavelet PDFs but this an interesting area to explore in the future 
\citep{davies_2016_fluct, becker_2018, kulkarni_2019, keating_2020, fahad_2020, gaikwad_2020_fluct}.

Implementing wavelet analysis can be challenging due to the large size of data vectors and covariance matrices involved. 
Often the cross correlations are ignored \citep{Lidz_2010, gaikwad_2020} which can lead to inaccurate parameter constraints. 
It is also impossible to remove the effect of both the resolution and the noise on the full wavelet amplitude PDFs, meaning you have to forward-model these effects, unlike in power spectrum analysis where the noise can be subtracted and the resolution effects removed by a window function correction.
When studying data, we would calculate the covariance matrix from bootstrapping the data itself as has been done when using other statistics of the Ly$\alpha$ forest (see \citet{Boera_2019} for example). 
This would reduce the computational time required since we would not be need to compute the covariance matrix for each model.

An interesting subject for future work will be to build an emulator using wavelet PDFs analogous to Ly$\alpha$ forest power spectrum emulators (see, e.g. \citet{walther_2019}). 
One issue of concern for wavelets is the large number of functions that need to be emulated (15 wavelets versus 1 power spectrum), although it could be possible to simply emulate the wavelet likelihood which is a single function.  
The emulation field has shifted towards iterative sampling, informed by the posterior probability distribution for a given observational dataset \citep{Rogers2019, Takhtaganov2021}, making emulating the likelihood consistent with current methods. 

In the future, our method of wavelet analysis can be applied to quasar data and more sophisticated simulations to obtain precision constraints on the thermal state of the IGM. 
One can expand our analysis to constrain the timing of reionization as well as models of dark matter. 
As mentioned above, our approach can also be adapted to analyses in proximity zones or searches for IGM temperature fluctuations exploiting the space-preserving properties of wavelets.


\section*{Acknowledgements}

We acknowledge helpful conversations with the ENIGMA group at UC Santa Barbara. 
JFH acknowledges support from the National Science Foundation under Grant No. 1816006.
We acknowledge PRACE for awarding us access to JUWELS at GCS@JSC, Germany.

\section*{Data Availability}

The simulation data analyzed in this article will be shared on reasonable request to the corresponding author.



\bibliographystyle{mnras}
\bibliography{overall} 

\begin{thebibliography}{}
\makeatletter
\relax
\def\mn@urlcharsother{\let\do\@makeother \do\$\do\&\do\#\do\^\do\_\do\%\do\~}
\def\mn@doi{\begingroup\mn@urlcharsother \@ifnextchar [ {\mn@doi@}
  {\mn@doi@[]}}
\def\mn@doi@[#1]#2{\def\@tempa{#1}\ifx\@tempa\@empty \href
  {http://dx.doi.org/#2} {doi:#2}\else \href {http://dx.doi.org/#2} {#1}\fi
  \endgroup}
\def\mn@eprint#1#2{\mn@eprint@#1:#2::\@nil}
\def\mn@eprint@arXiv#1{\href {http://arxiv.org/abs/#1} {{\tt arXiv:#1}}}
\def\mn@eprint@dblp#1{\href {http://dblp.uni-trier.de/rec/bibtex/#1.xml}
  {dblp:#1}}
\def\mn@eprint@#1:#2:#3:#4\@nil{\def\@tempa {#1}\def\@tempb {#2}\def\@tempc
  {#3}\ifx \@tempc \@empty \let \@tempc \@tempb \let \@tempb \@tempa \fi \ifx
  \@tempb \@empty \def\@tempb {arXiv}\fi \@ifundefined
  {mn@eprint@\@tempb}{\@tempb:\@tempc}{\expandafter \expandafter \csname
  mn@eprint@\@tempb\endcsname \expandafter{\@tempc}}}

\bibitem[\protect\citeauthoryear{{Almgren}, {Bell}, {Lijewski}, {Luki{\'c}}  \&
  {Van Andel}}{{Almgren} et~al.}{2013}]{almgren_2013}
{Almgren} A.~S.,  {Bell} J.~B.,  {Lijewski} M.~J.,  {Luki{\'c}} Z.,   {Van
  Andel} E.,  2013, \mn@doi [\apj] {10.1088/0004-637X/765/1/39}, \href
  {https://ui.adsabs.harvard.edu/abs/2013ApJ...765...39A} {765, 39}

\bibitem[\protect\citeauthoryear{{Becker}, {Rauch}  \& {Sargent}}{{Becker}
  et~al.}{2007}]{becker_2007}
{Becker} G.~D.,  {Rauch} M.,   {Sargent} W. L.~W.,  2007, \mn@doi [\apj]
  {10.1086/517866}, \href
  {https://ui.adsabs.harvard.edu/abs/2007ApJ...662...72B} {662, 72}

\bibitem[\protect\citeauthoryear{{Becker}, {Bolton}, {Haehnelt}  \&
  {Sargent}}{{Becker} et~al.}{2011}]{becker_2011}
{Becker} G.~D.,  {Bolton} J.~S.,  {Haehnelt} M.~G.,   {Sargent} W. L.~W.,
  2011, \mn@doi [\mnras] {10.1111/j.1365-2966.2010.17507.x}, \href
  {https://ui.adsabs.harvard.edu/abs/2011MNRAS.410.1096B} {410, 1096}

\bibitem[\protect\citeauthoryear{{Becker}, {Bolton}, {Madau}, {Pettini},
  {Ryan-Weber}  \& {Venemans}}{{Becker} et~al.}{2015}]{becker_2015}
{Becker} G.~D.,  {Bolton} J.~S.,  {Madau} P.,  {Pettini} M.,  {Ryan-Weber}
  E.~V.,   {Venemans} B.~P.,  2015, \mn@doi [\mnras] {10.1093/mnras/stu2646},
  \href {https://ui.adsabs.harvard.edu/abs/2015MNRAS.447.3402B} {447, 3402}

\bibitem[\protect\citeauthoryear{{Becker}, {Davies}, {Furlanetto}, {Malkan},
  {Boera}  \& {Douglass}}{{Becker} et~al.}{2018}]{becker_2018}
{Becker} G.~D.,  {Davies} F.~B.,  {Furlanetto} S.~R.,  {Malkan} M.~A.,  {Boera}
  E.,   {Douglass} C.,  2018, \mn@doi [\apj] {10.3847/1538-4357/aacc73}, \href
  {https://ui.adsabs.harvard.edu/abs/2018ApJ...863...92B} {863, 92}

\bibitem[\protect\citeauthoryear{Boera, Murphy, Becker  \& Bolton}{Boera
  et~al.}{2014}]{boera_2014}
Boera E.,  Murphy M.~T.,  Becker G.~D.,   Bolton J.~S.,  2014, \mn@doi [\mnras]
  {10.1093/mnras/stu660}, 441, 1916

\bibitem[\protect\citeauthoryear{Boera, Becker, Bolton  \& Nasir}{Boera
  et~al.}{2019}]{Boera_2019}
Boera E.,  Becker G.~D.,  Bolton J.~S.,   Nasir F.,  2019, \mn@doi [\apj]
  {10.3847/1538-4357/aafee4}, 872, 101

\bibitem[\protect\citeauthoryear{Bolton, Viel, Kim, Haehnelt  \&
  Carswell}{Bolton et~al.}{2008}]{bolton_2008}
Bolton J.~S.,  Viel M.,  Kim T.-S.,  Haehnelt M.~G.,   Carswell R.~F.,  2008,
  \mn@doi [\mnras] {10.1111/j.1365-2966.2008.13114.x}, 386, 1131

\bibitem[\protect\citeauthoryear{{Bolton}, {Becker}, {Wyithe}, {Haehnelt}  \&
  {Sargent}}{{Bolton} et~al.}{2010}]{bolton_2010}
{Bolton} J.~S.,  {Becker} G.~D.,  {Wyithe} J. S.~B.,  {Haehnelt} M.~G.,
  {Sargent} W. L.~W.,  2010, \mn@doi [\mnras]
  {10.1111/j.1365-2966.2010.16701.x}, \href
  {https://ui.adsabs.harvard.edu/abs/2010MNRAS.406..612B} {406, 612}

\bibitem[\protect\citeauthoryear{{Bolton}, {Becker}, {Raskutti}, {Wyithe},
  {Haehnelt}  \& {Sargent}}{{Bolton} et~al.}{2012}]{bolton_2012}
{Bolton} J.~S.,  {Becker} G.~D.,  {Raskutti} S.,  {Wyithe} J. S.~B.,
  {Haehnelt} M.~G.,   {Sargent} W. L.~W.,  2012, \mn@doi [\mnras]
  {10.1111/j.1365-2966.2011.19929.x}, \href
  {https://ui.adsabs.harvard.edu/abs/2012MNRAS.419.2880B} {419, 2880}

\bibitem[\protect\citeauthoryear{{Bolton}, {Becker}, {Haehnelt}  \&
  {Viel}}{{Bolton} et~al.}{2014}]{bolton_2014}
{Bolton} J.~S.,  {Becker} G.~D.,  {Haehnelt} M.~G.,   {Viel} M.,  2014, \mn@doi
  [\mnras] {10.1093/mnras/stt2374}, \href
  {https://ui.adsabs.harvard.edu/abs/2014MNRAS.438.2499B} {438, 2499}

\bibitem[\protect\citeauthoryear{{Bryan} \& {Machacek}}{{Bryan} \&
  {Machacek}}{2000}]{bryan_2000}
{Bryan} G.~L.,  {Machacek} M.~E.,  2000, \mn@doi [\apj] {10.1086/308735}, \href
  {https://ui.adsabs.harvard.edu/abs/2000ApJ...534...57B} {534, 57}

\bibitem[\protect\citeauthoryear{{Calura}, {Tescari}, {D'Odorico}, {Viel},
  {Cristiani}, {Kim}  \& {Bolton}}{{Calura} et~al.}{2012}]{calura_2012}
{Calura} F.,  {Tescari} E.,  {D'Odorico} V.,  {Viel} M.,  {Cristiani} S.,
  {Kim} T.~S.,   {Bolton} J.~S.,  2012, \mn@doi [\mnras]
  {10.1111/j.1365-2966.2012.20811.x}, \href
  {https://ui.adsabs.harvard.edu/abs/2012MNRAS.422.3019C} {422, 3019}

\bibitem[\protect\citeauthoryear{{D'Aloisio}, {McQuinn}, {Davies}  \&
  {Furlanetto}}{{D'Aloisio} et~al.}{2018}]{D_aloisio_2018}
{D'Aloisio} A.,  {McQuinn} M.,  {Davies} F.~B.,   {Furlanetto} S.~R.,  2018,
  \mn@doi [\mnras] {10.1093/mnras/stx2341}, \href
  {https://ui.adsabs.harvard.edu/abs/2018MNRAS.473..560D} {473, 560}

\bibitem[\protect\citeauthoryear{D'Aloisio, McQuinn, Maupin, Davies, Trac,
  Fuller  \& Sanderbeck}{D'Aloisio et~al.}{2019}]{D_Aloisio_2019}
D'Aloisio A.,  McQuinn M.,  Maupin O.,  Davies F.~B.,  Trac H.,  Fuller S.,
  Sanderbeck P. R.~U.,  2019, \mn@doi [\apj] {10.3847/1538-4357/ab0d83}, 874,
  154

\bibitem[\protect\citeauthoryear{{Davies} \& {Furlanetto}}{{Davies} \&
  {Furlanetto}}{2016}]{davies_2016_fluct}
{Davies} F.~B.,  {Furlanetto} S.~R.,  2016, \mn@doi [\mnras]
  {10.1093/mnras/stw931}, \href
  {https://ui.adsabs.harvard.edu/abs/2016MNRAS.460.1328D} {460, 1328}

\bibitem[\protect\citeauthoryear{Davies, Furlanetto  \& McQuinn}{Davies
  et~al.}{2016}]{davies_2016}
Davies F.~B.,  Furlanetto S.~R.,   McQuinn M.,  2016, \mn@doi [\mnras]
  {10.1093/mnras/stw055}, 457, 3006

\bibitem[\protect\citeauthoryear{Eilers, Hennawi  \& Lee}{Eilers
  et~al.}{2017}]{Eilers_2017}
Eilers A.-C.,  Hennawi J.~F.,   Lee K.-G.,  2017, \mn@doi [\apj]
  {10.3847/1538-4357/aa7e31}, 844, 136

\bibitem[\protect\citeauthoryear{{Fan} et~al.,}{{Fan} et~al.}{2006}]{fan_2006}
{Fan} X.,  et~al., 2006, \mn@doi [\aj] {10.1086/504836}, \href
  {https://ui.adsabs.harvard.edu/abs/2006AJ....132..117F} {132, 117}

\bibitem[\protect\citeauthoryear{{Fang} \& {White}}{{Fang} \&
  {White}}{2004}]{fang_2004}
{Fang} T.,  {White} M.,  2004, \mn@doi [\apjl] {10.1086/420965}, \href
  {https://ui.adsabs.harvard.edu/abs/2004ApJ...606L...9F} {606, L9}

\bibitem[\protect\citeauthoryear{{Gaikwad} et~al.,}{{Gaikwad}
  et~al.}{2020}]{gaikwad_2020_fluct}
{Gaikwad} P.,  et~al., 2020, \mn@doi [\mnras] {10.1093/mnras/staa907}, \href
  {https://ui.adsabs.harvard.edu/abs/2020MNRAS.494.5091G} {494, 5091}

\bibitem[\protect\citeauthoryear{{Gaikwad}, {Srianand}, {Haehnelt}  \&
  {Choudhury}}{{Gaikwad} et~al.}{2021}]{gaikwad_2020}
{Gaikwad} P.,  {Srianand} R.,  {Haehnelt} M.~G.,   {Choudhury} T.~R.,  2021,
  \mn@doi [\mnras] {10.1093/mnras/stab2017}, \href
  {https://ui.adsabs.harvard.edu/abs/2021MNRAS.506.4389G} {506, 4389}

\bibitem[\protect\citeauthoryear{{Garzilli}, {Bolton}, {Kim}, {Leach}  \&
  {Viel}}{{Garzilli} et~al.}{2012}]{garzilli_2012}
{Garzilli} A.,  {Bolton} J.~S.,  {Kim} T.~S.,  {Leach} S.,   {Viel} M.,  2012,
  \mn@doi [\mnras] {10.1111/j.1365-2966.2012.21223.x}, \href
  {https://ui.adsabs.harvard.edu/abs/2012MNRAS.424.1723G} {424, 1723}

\bibitem[\protect\citeauthoryear{{Garzilli}, {Boyarsky}  \&
  {Ruchayskiy}}{{Garzilli} et~al.}{2017}]{garzilli_2017}
{Garzilli} A.,  {Boyarsky} A.,   {Ruchayskiy} O.,  2017, \mn@doi [Physics
  Letters B] {10.1016/j.physletb.2017.08.022}, \href
  {https://ui.adsabs.harvard.edu/abs/2017PhLB..773..258G} {773, 258}

\bibitem[\protect\citeauthoryear{{Gnedin} \& {Hui}}{{Gnedin} \&
  {Hui}}{1998}]{gnedin_hui_1998}
{Gnedin} N.~Y.,  {Hui} L.,  1998, \mn@doi [\mnras]
  {10.1046/j.1365-8711.1998.01249.x}, \href
  {https://ui.adsabs.harvard.edu/abs/1998MNRAS.296...44G} {296, 44}

\bibitem[\protect\citeauthoryear{{Gunn} \& {Peterson}}{{Gunn} \&
  {Peterson}}{1965}]{gunn_peterson_1965}
{Gunn} J.~E.,  {Peterson} B.~A.,  1965, \mn@doi [\apj] {10.1086/148444}, \href
  {https://ui.adsabs.harvard.edu/abs/1965ApJ...142.1633G} {142, 1633}

\bibitem[\protect\citeauthoryear{{Haehnelt} \& {Steinmetz}}{{Haehnelt} \&
  {Steinmetz}}{1998}]{haehnelt_1998}
{Haehnelt} M.~G.,  {Steinmetz} M.,  1998, \mn@doi [\mnras]
  {10.1046/j.1365-8711.1998.01879.x}, \href
  {https://ui.adsabs.harvard.edu/abs/1998MNRAS.298L..21H} {298, L21}

\bibitem[\protect\citeauthoryear{Hiss, Walther, Hennawi, O{\~{n}}orbe, O'Meara,
  Rorai  \& Luki{\'{c}}}{Hiss et~al.}{2018}]{Hiss_2018}
Hiss H.,  Walther M.,  Hennawi J.~F.,  O{\~{n}}orbe J.,  O'Meara J.~M.,  Rorai
  A.,   Luki{\'{c}} Z.,  2018, \mn@doi [\apj] {10.3847/1538-4357/aada86}, 865,
  42

\bibitem[\protect\citeauthoryear{{Hui} \& {Gnedin}}{{Hui} \&
  {Gnedin}}{1997}]{hui_gnedin_1997}
{Hui} L.,  {Gnedin} N.~Y.,  1997, \mn@doi [\mnras] {10.1093/mnras/292.1.27},
  \href {https://ui.adsabs.harvard.edu/abs/1997MNRAS.292...27H} {292, 27}

\bibitem[\protect\citeauthoryear{{Hui} \& {Haiman}}{{Hui} \&
  {Haiman}}{2003}]{hui_haiman_2003}
{Hui} L.,  {Haiman} Z.,  2003, \mn@doi [\apj] {10.1086/377229}, \href
  {https://ui.adsabs.harvard.edu/abs/2003ApJ...596....9H} {596, 9}

\bibitem[\protect\citeauthoryear{Ir\ifmmode \check{s}\else
  \v{s}\fi{}i\ifmmode~\check{c}\else \v{c}\fi{} et~al.,}{Ir\ifmmode
  \check{s}\else \v{s}\fi{}i\ifmmode~\check{c}\else \v{c}\fi{}
  et~al.}{2017}]{Irsic_2017}
Ir\ifmmode \check{s}\else \v{s}\fi{}i\ifmmode~\check{c}\else \v{c}\fi{} V.,
  et~al., 2017, \mn@doi [Phys. Rev. D] {10.1103/PhysRevD.96.023522}, 96, 023522

\bibitem[\protect\citeauthoryear{{Keating}, {Weinberger}, {Kulkarni},
  {Haehnelt}, {Chardin}  \& {Aubert}}{{Keating} et~al.}{2020}]{keating_2020}
{Keating} L.~C.,  {Weinberger} L.~H.,  {Kulkarni} G.,  {Haehnelt} M.~G.,
  {Chardin} J.,   {Aubert} D.,  2020, \mn@doi [\mnras] {10.1093/mnras/stz3083},
  \href {https://ui.adsabs.harvard.edu/abs/2020MNRAS.491.1736K} {491, 1736}

\bibitem[\protect\citeauthoryear{Khrykin, Hennawi, McQuinn  \& Worseck}{Khrykin
  et~al.}{2016}]{Khrykin_2016}
Khrykin I.~S.,  Hennawi J.~F.,  McQuinn M.,   Worseck G.,  2016, \mn@doi [\apj]
  {10.3847/0004-637x/824/2/133}, 824, 133

\bibitem[\protect\citeauthoryear{{Kulkarni}, {Hennawi}, {O{\~n}orbe}, {Rorai}
  \& {Springel}}{{Kulkarni} et~al.}{2015}]{kulkarni_2015}
{Kulkarni} G.,  {Hennawi} J.~F.,  {O{\~n}orbe} J.,  {Rorai} A.,   {Springel}
  V.,  2015, \mn@doi [\apj] {10.1088/0004-637X/812/1/30}, \href
  {https://ui.adsabs.harvard.edu/abs/2015ApJ...812...30K} {812, 30}

\bibitem[\protect\citeauthoryear{{Kulkarni}, {Keating}, {Haehnelt}, {Bosman},
  {Puchwein}, {Chardin}  \& {Aubert}}{{Kulkarni} et~al.}{2019}]{kulkarni_2019}
{Kulkarni} G.,  {Keating} L.~C.,  {Haehnelt} M.~G.,  {Bosman} S. E.~I.,
  {Puchwein} E.,  {Chardin} J.,   {Aubert} D.,  2019, \mn@doi [\mnras]
  {10.1093/mnrasl/slz025}, \href
  {https://ui.adsabs.harvard.edu/abs/2019MNRAS.485L..24K} {485, L24}

\bibitem[\protect\citeauthoryear{Lai, Lidz, Hernquist  \& Zaldarriaga}{Lai
  et~al.}{2006}]{Lai_2006}
Lai K.,  Lidz A.,  Hernquist L.,   Zaldarriaga M.,  2006, \mn@doi [\apj]
  {10.1086/503320}, 644, 61

\bibitem[\protect\citeauthoryear{Lee et~al.,}{Lee et~al.}{2015}]{Lee_2015}
Lee K.-G.,  et~al., 2015, \mn@doi [\apj] {10.1088/0004-637x/799/2/196}, 799,
  196

\bibitem[\protect\citeauthoryear{{Lidz} \& {Malloy}}{{Lidz} \&
  {Malloy}}{2014}]{lidz_2014}
{Lidz} A.,  {Malloy} M.,  2014, \mn@doi [\apj] {10.1088/0004-637X/788/2/175},
  \href {https://ui.adsabs.harvard.edu/abs/2014ApJ...788..175L} {788, 175}

\bibitem[\protect\citeauthoryear{Lidz, Heitmann, Hui, Habib, Rauch  \&
  Sargent}{Lidz et~al.}{2006}]{Lidz_2006}
Lidz A.,  Heitmann K.,  Hui L.,  Habib S.,  Rauch M.,   Sargent W. L.~W.,
  2006, \mn@doi [\apj] {10.1086/498699}, 638, 27–44

\bibitem[\protect\citeauthoryear{Lidz, Faucher-Gigu{\`{e}}re, Dall'Aglio,
  McQuinn, Fechner, Zaldarriaga, Hernquist  \& Dutta}{Lidz
  et~al.}{2010}]{Lidz_2010}
Lidz A.,  Faucher-Gigu{\`{e}}re C.-A.,  Dall'Aglio A.,  McQuinn M.,  Fechner
  C.,  Zaldarriaga M.,  Hernquist L.,   Dutta S.,  2010, \mn@doi [\apj]
  {10.1088/0004-637x/718/1/199}, 718, 199

\bibitem[\protect\citeauthoryear{{Luki{\'c}}, {Stark}, {Nugent}, {White},
  {Meiksin}  \& {Almgren}}{{Luki{\'c}} et~al.}{2015}]{Lukic_2015}
{Luki{\'c}} Z.,  {Stark} C.~W.,  {Nugent} P.,  {White} M.,  {Meiksin} A.~A.,
  {Almgren} A.,  2015, \mn@doi [\mnras] {10.1093/mnras/stu2377}, \href
  {https://ui.adsabs.harvard.edu/abs/2015MNRAS.446.3697L} {446, 3697}

\bibitem[\protect\citeauthoryear{{Lynds}}{{Lynds}}{1971}]{lynds_1971}
{Lynds} R.,  1971, \mn@doi [\apjl] {10.1086/180695}, \href
  {https://ui.adsabs.harvard.edu/abs/1971ApJ...164L..73L} {164, L73}

\bibitem[\protect\citeauthoryear{{McDonald}, {Miralda-Escud{\'e}}, {Rauch},
  {Sargent}, {Barlow}  \& {Cen}}{{McDonald} et~al.}{2001}]{mcdonald_2001}
{McDonald} P.,  {Miralda-Escud{\'e}} J.,  {Rauch} M.,  {Sargent} W. L.~W.,
  {Barlow} T.~A.,   {Cen} R.,  2001, \mn@doi [\apj] {10.1086/323426}, \href
  {https://ui.adsabs.harvard.edu/abs/2001ApJ...562...52M} {562, 52}

\bibitem[\protect\citeauthoryear{McQuinn}{McQuinn}{2012}]{mcquinn_2012}
McQuinn M.,  2012, \mn@doi [\mnras] {10.1111/j.1365-2966.2012.21792.x}, 426,
  1349

\bibitem[\protect\citeauthoryear{{McQuinn} \& {Upton Sanderbeck}}{{McQuinn} \&
  {Upton Sanderbeck}}{2016}]{Mcquinn_upton_2016}
{McQuinn} M.,  {Upton Sanderbeck} P.~R.,  2016, \mn@doi [\mnras]
  {10.1093/mnras/stv2675}, \href
  {https://ui.adsabs.harvard.edu/abs/2016MNRAS.456...47M} {456, 47}

\bibitem[\protect\citeauthoryear{McQuinn, Hernquist, Lidz  \&
  Zaldarriaga}{McQuinn et~al.}{2011}]{McQuinn_2011}
McQuinn M.,  Hernquist L.,  Lidz A.,   Zaldarriaga M.,  2011, \mn@doi [\mnras]
  {10.1111/j.1365-2966.2011.18788.x}, 415, 977

\bibitem[\protect\citeauthoryear{{Meiksin}}{{Meiksin}}{2000}]{meiksin_2000}
{Meiksin} A.,  2000, \mn@doi [\mnras] {10.1046/j.1365-8711.2000.03315.x}, \href
  {https://ui.adsabs.harvard.edu/abs/2000MNRAS.314..566M} {314, 566}

\bibitem[\protect\citeauthoryear{Miralda-Escudé \& Rees}{Miralda-Escudé \&
  Rees}{1994}]{miralda_escude_1994}
Miralda-Escudé J.,  Rees M.~J.,  1994, \mn@doi [\mnras]
  {10.1093/mnras/266.2.343}, 266, 343

\bibitem[\protect\citeauthoryear{Narayanan, Spergel, Davé  \& Ma}{Narayanan
  et~al.}{2000}]{Narayanan_2000}
Narayanan V.~K.,  Spergel D.~N.,  Davé R.,   Ma C.-P.,  2000, \mn@doi [\apj]
  {10.1086/317269}, 543, L103–L106

\bibitem[\protect\citeauthoryear{{Nasir} \& {D'Aloisio}}{{Nasir} \&
  {D'Aloisio}}{2020}]{fahad_2020}
{Nasir} F.,  {D'Aloisio} A.,  2020, \mn@doi [\mnras] {10.1093/mnras/staa894},
  \href {https://ui.adsabs.harvard.edu/abs/2020MNRAS.494.3080N} {494, 3080}

\bibitem[\protect\citeauthoryear{{Nasir}, {Bolton}  \& {Becker}}{{Nasir}
  et~al.}{2016}]{Nasir_2016}
{Nasir} F.,  {Bolton} J.~S.,   {Becker} G.~D.,  2016, \mn@doi [\mnras]
  {10.1093/mnras/stw2147}, \href
  {https://ui.adsabs.harvard.edu/abs/2016MNRAS.463.2335N} {463, 2335}

\bibitem[\protect\citeauthoryear{{O{\~n}orbe}, {Hennawi}  \&
  {Luki{\'c}}}{{O{\~n}orbe} et~al.}{2017}]{onorbe_2017}
{O{\~n}orbe} J.,  {Hennawi} J.~F.,   {Luki{\'c}} Z.,  2017, \mn@doi [\apj]
  {10.3847/1538-4357/aa6031}, \href
  {https://ui.adsabs.harvard.edu/abs/2017ApJ...837..106O} {837, 106}

\bibitem[\protect\citeauthoryear{O{\~{n}}orbe, Hennawi, Luki{\'{c}}  \&
  Walther}{O{\~{n}}orbe et~al.}{2017}]{Onorbe_2017_planck}
O{\~{n}}orbe J.,  Hennawi J.~F.,  Luki{\'{c}} Z.,   Walther M.,  2017, \mn@doi
  [\apj] {10.3847/1538-4357/aa898d}, 847, 63

\bibitem[\protect\citeauthoryear{O{\~{n}}orbe, Davies, Lukić, Hennawi  \&
  Sorini}{O{\~{n}}orbe et~al.}{2019}]{Onorbe_2019}
O{\~{n}}orbe J.,  Davies F.~B.,  Lukić Z.,  Hennawi J.~F.,   Sorini D.,  2019,
  \mn@doi [\mnras] {10.1093/mnras/stz984}, 486, 4075

\bibitem[\protect\citeauthoryear{{Planck Collaboration} et~al.,}{{Planck
  Collaboration} et~al.}{2020}]{Planck_2018}
{Planck Collaboration} et~al., 2020, \mn@doi [A\&A]
  {10.1051/0004-6361/201833910}, 641, A6

\bibitem[\protect\citeauthoryear{{Puchwein}, {Bolton}, {Haehnelt}, {Madau},
  {Becker}  \& {Haardt}}{{Puchwein} et~al.}{2015}]{puchwein_2015}
{Puchwein} E.,  {Bolton} J.~S.,  {Haehnelt} M.~G.,  {Madau} P.,  {Becker}
  G.~D.,   {Haardt} F.,  2015, \mn@doi [\mnras] {10.1093/mnras/stv773}, \href
  {https://ui.adsabs.harvard.edu/abs/2015MNRAS.450.4081P} {450, 4081}

\bibitem[\protect\citeauthoryear{Rauch}{Rauch}{1998}]{Rauch_1998}
Rauch M.,  1998, \mn@doi [ARA&A] {10.1146/annurev.astro.36.1.267}, 36,
  267–316

\bibitem[\protect\citeauthoryear{{Ricotti}, {Gnedin}  \& {Shull}}{{Ricotti}
  et~al.}{2000}]{ricotti_2000}
{Ricotti} M.,  {Gnedin} N.~Y.,   {Shull} J.~M.,  2000, \mn@doi [\apj]
  {10.1086/308733}, \href
  {https://ui.adsabs.harvard.edu/abs/2000ApJ...534...41R} {534, 41}

\bibitem[\protect\citeauthoryear{{Rogers}, {Peiris}, {Pontzen}, {Bird}, {Verde}
   \& {Font-Ribera}}{{Rogers} et~al.}{2019}]{Rogers2019}
{Rogers} K.~K.,  {Peiris} H.~V.,  {Pontzen} A.,  {Bird} S.,  {Verde} L.,
  {Font-Ribera} A.,  2019, \mn@doi [\jcap] {10.1088/1475-7516/2019/02/031},
  \href {https://ui.adsabs.harvard.edu/abs/2019JCAP...02..031R} {2019, 031}

\bibitem[\protect\citeauthoryear{Rorai et~al.,}{Rorai
  et~al.}{2017}]{Rorai_2017}
Rorai A.,  et~al., 2017, \mn@doi [Science] {10.1126/science.aaf9346}, 356, 418

\bibitem[\protect\citeauthoryear{{Rorai}, {Carswell}, {Haehnelt}, {Becker},
  {Bolton}  \& {Murphy}}{{Rorai} et~al.}{2018}]{rorai_2018}
{Rorai} A.,  {Carswell} R.~F.,  {Haehnelt} M.~G.,  {Becker} G.~D.,  {Bolton}
  J.~S.,   {Murphy} M.~T.,  2018, \mn@doi [\mnras] {10.1093/mnras/stx2862},
  \href {https://ui.adsabs.harvard.edu/abs/2018MNRAS.474.2871R} {474, 2871}

\bibitem[\protect\citeauthoryear{{Rudie}, {Steidel}  \& {Pettini}}{{Rudie}
  et~al.}{2012}]{rudie_2012}
{Rudie} G.~C.,  {Steidel} C.~C.,   {Pettini} M.,  2012, \mn@doi [\apjl]
  {10.1088/2041-8205/757/2/L30}, \href
  {https://ui.adsabs.harvard.edu/abs/2012ApJ...757L..30R} {757, L30}

\bibitem[\protect\citeauthoryear{{Schaye}, {Theuns}, {Leonard}  \&
  {Efstathiou}}{{Schaye} et~al.}{1999}]{schaye_1999}
{Schaye} J.,  {Theuns} T.,  {Leonard} A.,   {Efstathiou} G.,  1999, \mn@doi
  [\mnras] {10.1046/j.1365-8711.1999.02956.x}, \href
  {https://ui.adsabs.harvard.edu/abs/1999MNRAS.310...57S} {310, 57}

\bibitem[\protect\citeauthoryear{{Schaye}, {Theuns}, {Rauch}, {Efstathiou}  \&
  {Sargent}}{{Schaye} et~al.}{2000}]{schaye_2000}
{Schaye} J.,  {Theuns} T.,  {Rauch} M.,  {Efstathiou} G.,   {Sargent} W. L.~W.,
   2000, \mn@doi [\mnras] {10.1046/j.1365-8711.2000.03815.x}, \href
  {https://ui.adsabs.harvard.edu/abs/2000MNRAS.318..817S} {318, 817}

\bibitem[\protect\citeauthoryear{{Takhtaganov}, {Luki{\'c}}, {M{\"u}ller}  \&
  {Morozov}}{{Takhtaganov} et~al.}{2021}]{Takhtaganov2021}
{Takhtaganov} T.,  {Luki{\'c}} Z.,  {M{\"u}ller} J.,   {Morozov} D.,  2021,
  \mn@doi [\apj] {10.3847/1538-4357/abc8ed}, \href
  {https://ui.adsabs.harvard.edu/abs/2021ApJ...906...74T} {906, 74}

\bibitem[\protect\citeauthoryear{{Theuns} \& {Zaroubi}}{{Theuns} \&
  {Zaroubi}}{2000}]{theuns_zaroubi_2000}
{Theuns} T.,  {Zaroubi} S.,  2000, \mn@doi [\mnras]
  {10.1046/j.1365-8711.2000.03729.x}, \href
  {https://ui.adsabs.harvard.edu/abs/2000MNRAS.317..989T} {317, 989}

\bibitem[\protect\citeauthoryear{{Theuns}, {Schaye}  \& {Haehnelt}}{{Theuns}
  et~al.}{2000}]{theuns_2000}
{Theuns} T.,  {Schaye} J.,   {Haehnelt} M.~G.,  2000, \mn@doi [\mnras]
  {10.1046/j.1365-8711.2000.03423.x}, \href
  {https://ui.adsabs.harvard.edu/abs/2000MNRAS.315..600T} {315, 600}

\bibitem[\protect\citeauthoryear{Theuns, Zaroubi, Kim, Tzanavaris  \&
  Carswell}{Theuns et~al.}{2002a}]{theuns_2002_fluct}
Theuns T.,  Zaroubi S.,  Kim T.-S.,  Tzanavaris P.,   Carswell R.~F.,  2002a,
  \mn@doi [\mnras] {10.1046/j.1365-8711.2002.05316.x}, 332, 367

\bibitem[\protect\citeauthoryear{{Theuns}, {Schaye}, {Zaroubi}, {Kim},
  {Tzanavaris}  \& {Carswell}}{{Theuns} et~al.}{2002b}]{theuns_2002_wavelet}
{Theuns} T.,  {Schaye} J.,  {Zaroubi} S.,  {Kim} T.-S.,  {Tzanavaris} P.,
  {Carswell} B.,  2002b, \mn@doi [\apjl] {10.1086/339998}, \href
  {https://ui.adsabs.harvard.edu/abs/2002ApJ...567L.103T} {567, L103}

\bibitem[\protect\citeauthoryear{Viel, Bolton  \& Haehnelt}{Viel
  et~al.}{2009}]{viel_2009}
Viel M.,  Bolton J.~S.,   Haehnelt M.~G.,  2009, \mn@doi [\mnras: Letters]
  {10.1111/j.1745-3933.2009.00720.x}, 399, L39

\bibitem[\protect\citeauthoryear{Viel, Becker, Bolton  \& Haehnelt}{Viel
  et~al.}{2013}]{Viel_2013}
Viel M.,  Becker G.~D.,  Bolton J.~S.,   Haehnelt M.~G.,  2013, \mn@doi [Phys.
  Rev. D] {10.1103/physrevd.88.043502}, 88

\bibitem[\protect\citeauthoryear{Walther, Hennawi, Hiss, O{\~{n}}orbe, Lee,
  Rorai  \& O'Meara}{Walther et~al.}{2017}]{Walther_2017}
Walther M.,  Hennawi J.~F.,  Hiss H.,  O{\~{n}}orbe J.,  Lee K.-G.,  Rorai A.,
   O'Meara J.,  2017, \mn@doi [\apj] {10.3847/1538-4357/aa9c81}, 852, 22

\bibitem[\protect\citeauthoryear{{Walther}, {O{\~n}orbe}, {Hennawi}  \&
  {Luki{\'c}}}{{Walther} et~al.}{2019}]{walther_2019}
{Walther} M.,  {O{\~n}orbe} J.,  {Hennawi} J.~F.,   {Luki{\'c}} Z.,  2019,
  \mn@doi [\apj] {10.3847/1538-4357/aafad1}, \href
  {https://ui.adsabs.harvard.edu/abs/2019ApJ...872...13W} {872, 13}

\bibitem[\protect\citeauthoryear{{Y{\`e}che}, {Palanque-Delabrouille}, {Baur}
  \& {du Mas des Bourboux}}{{Y{\`e}che} et~al.}{2017}]{yeche_2017}
{Y{\`e}che} C.,  {Palanque-Delabrouille} N.,  {Baur} J.,   {du Mas des
  Bourboux} H.,  2017, \mn@doi [\jcap] {10.1088/1475-7516/2017/06/047}, \href
  {https://ui.adsabs.harvard.edu/abs/2017JCAP...06..047Y} {2017, 047}

\bibitem[\protect\citeauthoryear{{Zaldarriaga}}{{Zaldarriaga}}{2002}]{Zaldarriaga_2002}
{Zaldarriaga} M.,  2002, \mn@doi [\apj] {10.1086/324212}, \href
  {https://ui.adsabs.harvard.edu/abs/2002ApJ...564..153Z} {564, 153}

\bibitem[\protect\citeauthoryear{{Zaldarriaga}, {Hui}  \&
  {Tegmark}}{{Zaldarriaga} et~al.}{2001}]{Zaldarriaga_2001}
{Zaldarriaga} M.,  {Hui} L.,   {Tegmark} M.,  2001, \mn@doi [\apj]
  {10.1086/321652}, \href
  {https://ui.adsabs.harvard.edu/abs/2001ApJ...557..519Z} {557, 519}

\makeatother
\end{thebibliography}



\appendix

\section{Likelihood Choice} \label{section:appendix likelihood choice}

We chose to use a multivariate Gaussian distribution for the likelihood of our data. 
This assumption explicitly means that if we take multiple mock data sets and look at the distribution of a single point in our data vector (either a single bin from the wavelet amplitude histogram or a single $k$ band from power spectra) it will be Gaussian. 
It also assumes that when looking at any two points from the data vector, we expect the resulting distribution to be a two-dimensional Gaussian (thus looking at many points will be a multi-dimensional Gaussian). 

\begin{figure}
	\includegraphics[width=\columnwidth]{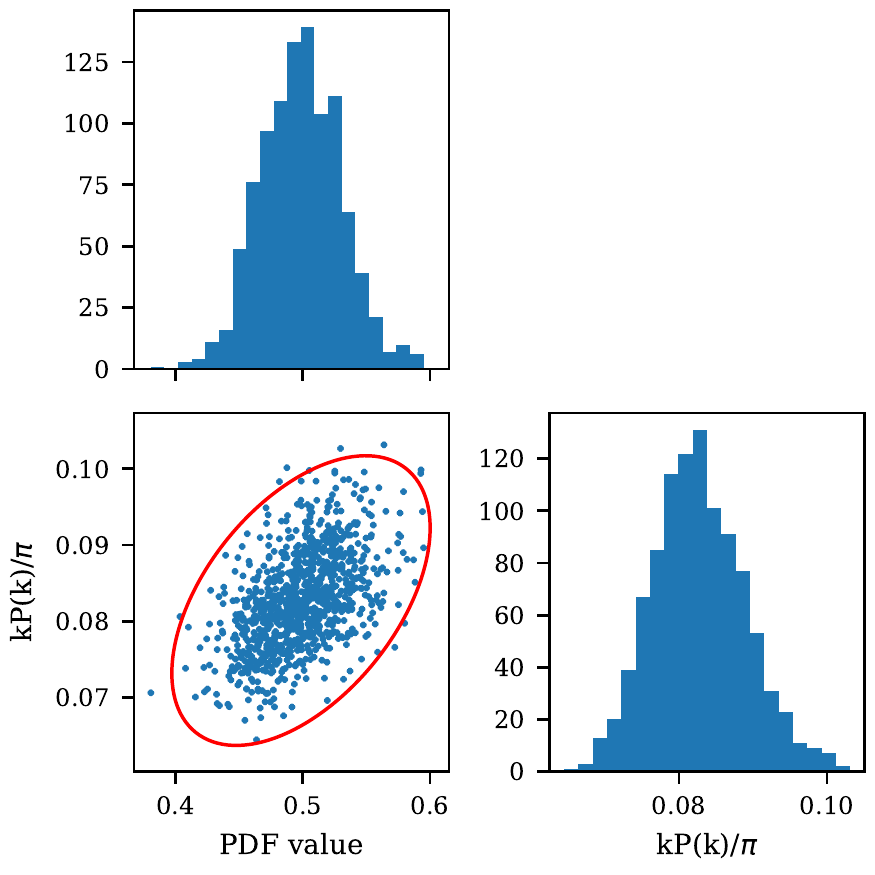}
    \caption{
        The distribution of mock draws for the power spectrum at $k = \SI{0.12}{\second\per\kilo\meter}$ and the wavelet amplitude PDF at $A_n = 0.32$. 
        The bottom left panel shows the 2D distribution of these bins where the red ellipse shows the 3$\sigma$ region calculated from the covariance matrix for these two bins. 
        The bottom right panel shows the distribution of values only for the power spectrum. 
        The top left panel shows the distribution of values only for the wavelet amplitude PDF. 
        All panels show good agreement with the assumption of a multi-variate Gaussian distribution.
        }
    \label{fig:multi_gaus_corner}
\end{figure}

We can visually check the assumption that any two points from the data vector will result in a two-dimensional Gaussian distribution over many mocks. 
We will show this for one point in the power spectrum ($k = \SI{0.12}{\second\per\kilo\meter}$) and one point in the wavelet amplitude PDF for $s_n = \SI{51.09}{\kilo\meter\per\second}$ ($A_n = 0.32$). 
The distribution of these values for 1,000 mock draws is shown in Figure \ref{fig:multi_gaus_corner} along with the distributions of the individual Histogram and $P_{\rm F}(k)$ values. 
The red ellipse represents the $3\sigma$ contour from the $2 \times 2$ covariance matrix calculated for these two bins. 
The two dimensional distribution agrees very well with the ellipse by eye and the two histograms also visually appear Gaussian within the errors expected from counting statistics. 

\section{Covariance Matrix Calculation}\label{section:appendix covariance smoothing}

\begin{figure}
	\includegraphics[width=\columnwidth]{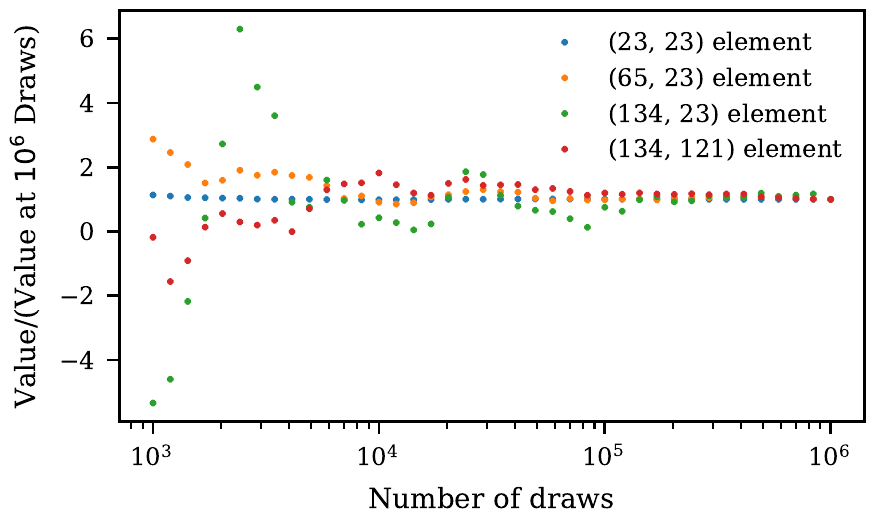}
    \caption{
        This figure shows the values of four distinct elements in the covariance matrix for the wavelet amplitude PDFs as we increased the number of mock draws used to calculate the covariance matrix.
        For simplicity, these distinct values are labeled by their index, rather than the smoothing scale and wavelet amplitude values associated for each bin in the PDFs. 
        These points were chosen such that we had one on the diagonal, one off the diagonal where there are strong correlations, and two off the diagonal where there are weak correlations.
        As we approach $10^6$ draws the values converge, showing that $10^6$ is sufficient for our covariance matrix calculation. 
        }
    \label{fig:covariance_convergence}
\end{figure}

As defined by equation \eqref{eq:covariance}, each model covariance matrix is calculated from mock draws of the data.
This will inherently be a noisy calculation that will converge as $1 / \sqrt{N}$ where $N$ is the number of draws in the covariance matrix. 
As stated in the text, we used 1,000,000 mock draws when calculating the covariance matrix for the wavelet amplitude PDFs as well as the combination of the wavelet amplitude PDFs and the power spectrum. 
To check that 1,000,000 mock draws are sufficient to minimize the error from this calculation, we looked at the behavior of elements of the wavelet amplitude PDF covariance matrix in Figure \ref{fig:covariance_convergence}. 
The values in the plot have been normalized such that at $10^6$ draws they are 1. 
The four elements have been chosen such that there is one on the diagonal, one off the diagonal where there are strong correlations, and two off the diagonal where there are weak correlations.
This plot shows that as we approach $10^6$ draws the values vary significantly less than they do at lower values and thus the covariance elements are converging.  

For both the wavelet amplitude PDFs and the combined power spectrum and wavelet amplitude PDFs covariance matrix, the data vector is long enough that there are many elements with very small cross correlations. 
These small values vary more (as seen in the (134, 23) and (134, 121) elements in Figure \ref{fig:covariance_convergence}) such that they can still have non-negligible noise. 
This noise in the covariance leads to noise in the posterior measurement on $T_0$ (as discussed in Sections \ref{section:amp likelihood} and \ref{section:combined likelihood}). 
At this point, for computational reasons, we decided not to increase the number of mock draws of data.
Instead, we chose to smooth the covariance matrix across our thermal grid with a spline. 
We did this by fitting each individual element of the covariance matrix to spline with 20 equally spaced breakpoints.
This results in $150 \times 150$ ($165 \times 165$) splines for the wavelet amplitude PDFs (combined) covariance matrix. 
We chose 20 breakpoints to allow the spline to be flexible enough to find real patterns in the data while smoothing out noise. 

\begin{figure}
	\includegraphics[width=\columnwidth]{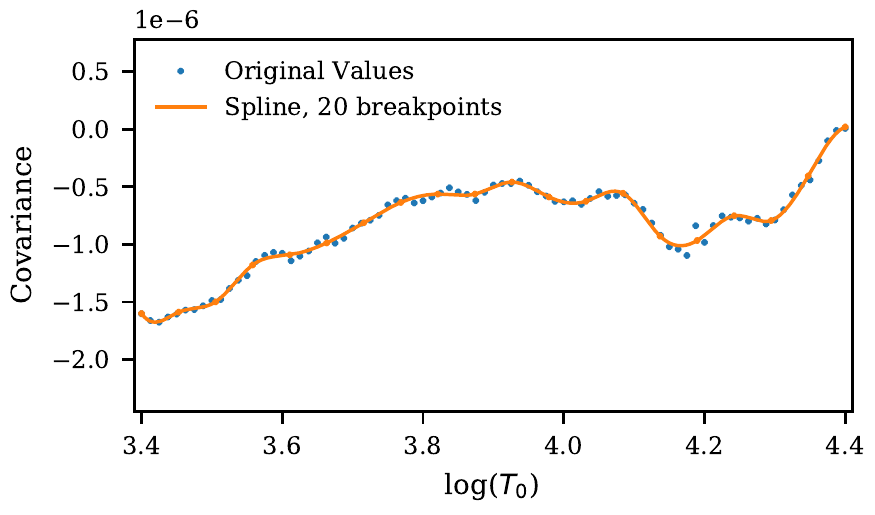}
    \caption{
        The covariance value for the same bin of the covariance matrix at different $T_0$. 
        The bin chosen here corresponds to the histogram bin of $A_n = 0.027$ for $s_n = \SI{51.09}{\kilo\meter\per\second}$ and the power spectrum band $k = \SI{0.41}{\second\per\kilo\meter}$) and $A_n = 0.027$. 
        The solid line shows the spline fit to these points with 20 equally spaced breakpoints.
        }
    \label{fig:covariance_smoothing}
\end{figure}

We show one example of this spline in Figure \ref{fig:covariance_smoothing}. 
This is the spline for one of the elements of the covariance matrix shown in Figures \ref{fig:combined_correlation_s9} and \ref{fig:combined_correlation}. 
Specifically, this is for $k = \SI{0.41}{\second\per\kilo\meter}$) and $A_n = 0.027$. 
This figure shows how the spline can replicate patterns in the calculated covariance matrix values while still reducing noise. 
For example, with the noisy values near $\log(T_0) = 4.2$.

The covariance matrices used in equation \eqref{eq:gauss_like} for both the wavelet analysis and the combined wavelet and power spectrum analysis use the values of the spline at every $T_0$.

\section{Redshift 6} \label{sec:redshift_6}

Here we have included the figures for a mock data set at $z = 6$. 
As a reminder, at $z=6$ we considered $\langle F \rangle = 0.011$, $R=30,000$, and SNR $= 35$. 
We restricted our study to a higher SNR because the lower observed flux makes the noise power more dominant, see equation \eqref{eq:noise power} for details. 
The mock data set still considers 8 quasars (equivalent to 29 simulation skewers). 

\begin{figure}
	\includegraphics[width=\columnwidth]{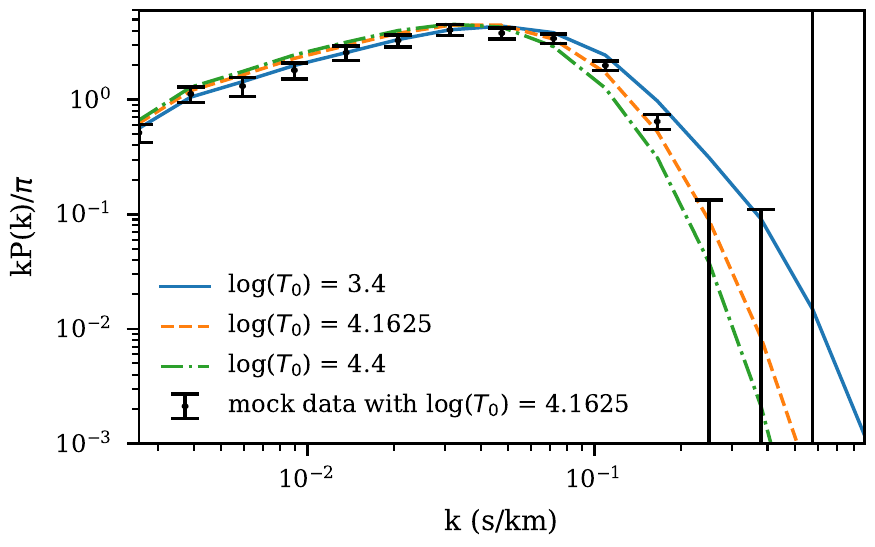}
    \caption{
        The power spectrum measurement, $P_{\rm F}(k)$, for one mock data set with $\log(T_0) = 4.1625$ (black points) at $z=6$. 
        The $1\sigma$ error bars are calculated from the square root of the diagonal of the covariance matrix.
        Also shown are model values of the power spectra for three different values of $T_0$: $\log(T_0) = 3.4$ (blue), $\log(T_0) = 4.1625$ (orange), and $\log(T_0) = 4.4$ (green). 
    }
    \label{fig:z6_power_mock}
\end{figure}

\begin{figure}
	\includegraphics[width=\columnwidth]{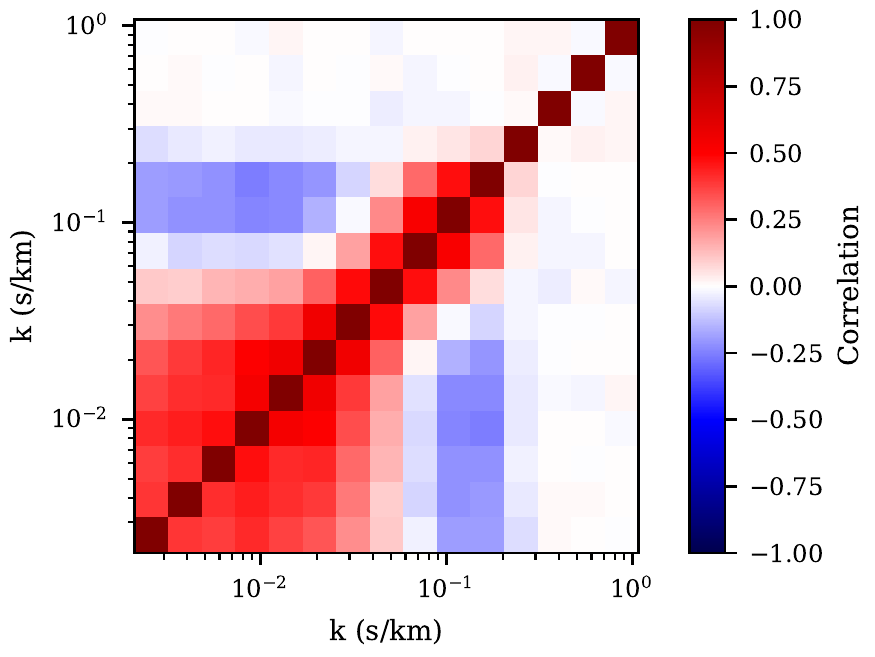}
    \caption{
        The correlation matrix for the power spectrum at $\log(T_0) = 4.1625$ and $z=6$.
        The very weak correlations seen in the regions where $k > \SI{0.2}{\second\per\kilo\meter}$ are due to uncorrelated random Gaussian noise which dominates the signal on small-scales (high $k$). 
    }
    \label{fig:z6_power_correlation}
\end{figure}

First, in Figure \ref{fig:z6_power_mock} we show a mock power spectrum at $\log(T_0) = 4.1625$ with three models analogous to Figure \ref{fig:one_mock_power} at $z=5$. 
The error bars blow up at the largest values of $k$ due to removing the noise power which is dominant at these small scales as well as correcting the window function. 
This Figure also shows three model power spectra at $z=6$: $\log(T_0) = 3.4$ (blue), $\log(T_0) = 4.1625$ (orange), and $\log(T_0) = 4.4$ (green). 
The corresponding correlation plot for $\log(T_0) = 4.1625$ is shown in Figure \ref{fig:z6_power_correlation}. 
The structure here looks quite different than the one at $z=5$ shown in Figure \ref{fig:power_correlation}.
The correlations come from underlying correlations in the high-$z$ Ly$\alpha$ forest with the overall low mean flux.
Again, the off-diagonal elements for high-$k$ (small scales) are very small due to the dominance of the noise power over the signal at these scales. 

\begin{figure*}
	\includegraphics[width=\columnwidth*2]{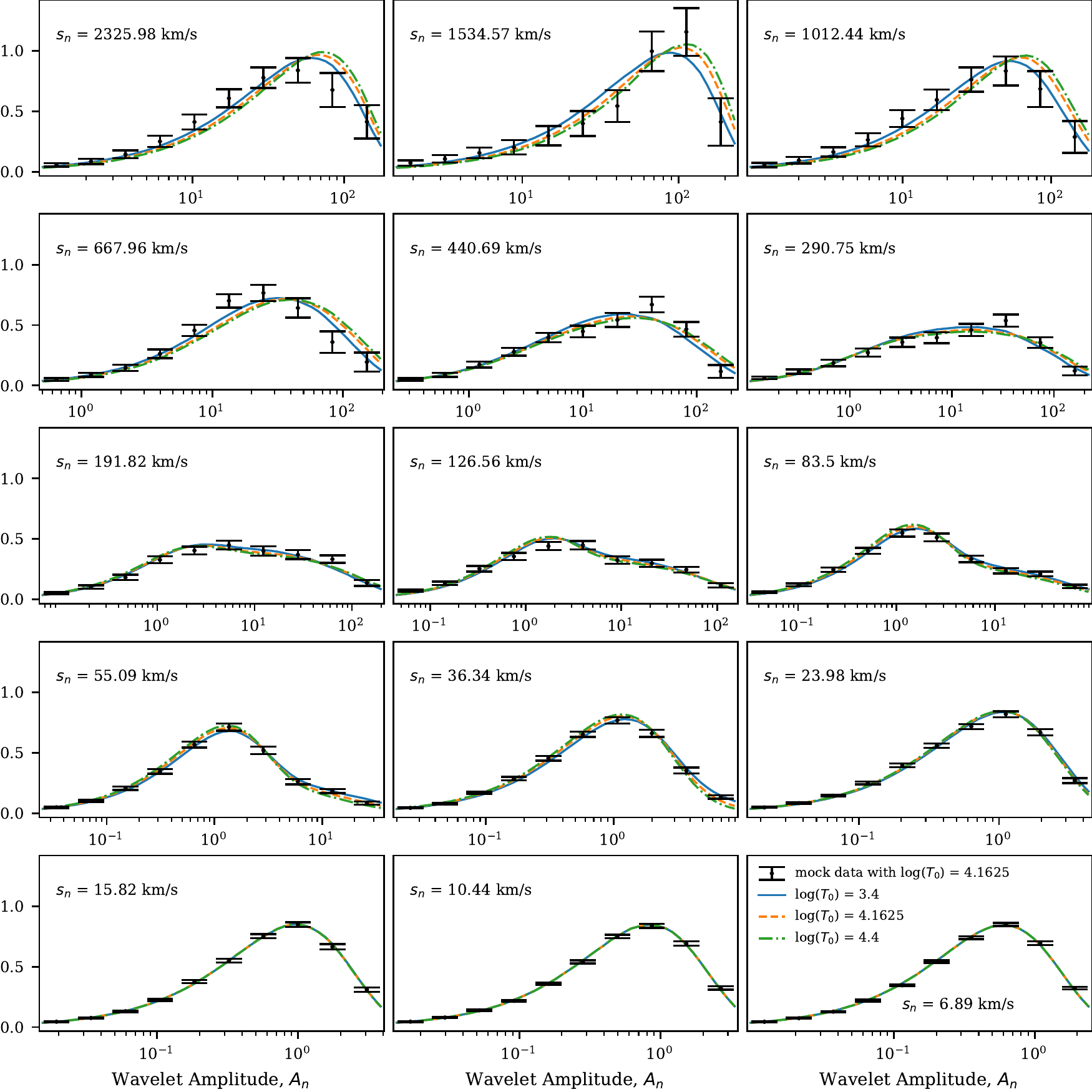}
    \caption{
        The black points show the PDFs from one mock data set for each $s_n$ with $\log(T_0) = 4.1265$ and $z=6$.
        The $1\sigma$ error bars are calculated from the square root of the diagonal of the covariance matrix.
        Each panel also shows the "model" values of the PDFs from the stated smoothing scale for three different values of $T_0$: $\log(T_0) = 3.4$ (blue), $\log(T_0) = 4.1625$ (orange), and $\log(T_0) = 4.4$ (green).
        This redshift shows broad PDFs with arguable two bumps in the mid-range values of $s_n$ where the flux and noise power levels are comparable.
    }
    \label{fig:z6_amps_mock}
\end{figure*}

\begin{figure*}
	\includegraphics[width=\columnwidth*2]{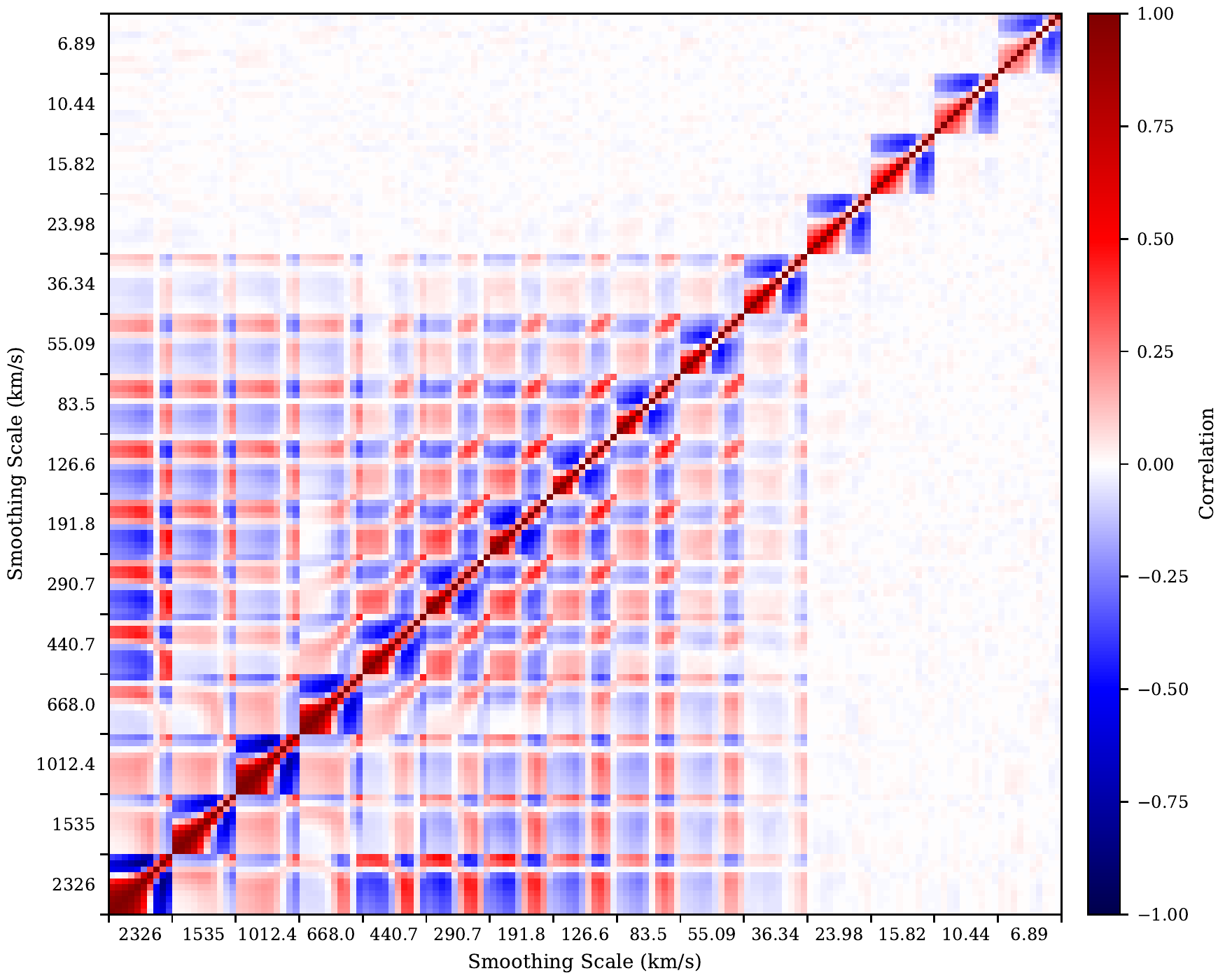}
    \caption{
        The correlation matrix for fifteen wavelet amplitude PDFs at $\log(T_0) = 4.1625$ and $z=6$.
        The wavelet amplitude PDFs for large smoothing scales, $\SI{2326}{\kilo\meter\per\second} \geq s_n \geq \SI{36.34}{\kilo\meter\per\second}$, have significant correlations off the diagonal.
        The off diagonal blocks show a similar repeating shape to those on the diagonal modified by numbers. 
        These numbers are positive close to the diagonal and are negative further from the diagonal. 
        For $s_n \leq \SI{29.98}{\kilo\meter\per\second}$, the wavelet amplitudes begin to be dominated by noise, so the correlations between PDFs for different values of $s_n$ become very small. 
        This pattern mimics that seen in the power spectrum correlation shown in Figure \ref{fig:z6_power_correlation}.
        The pattern of the diagonal blocks are the most different for the mid range values of $\SI{290.7}{\kilo\meter\per\second} \geq s_n \geq \SI{83.5}{\kilo\meter\per\second}$, which is where the PDF shapes are the most different from the typical shape, as seen in Figure \ref{fig:z6_amps_mock}.
    }
    \label{fig:z6_amps_correlation}
\end{figure*}

Next, we look at the figures relevant for the wavelet amplitude PDFs at $z=6$. 
The mock wavelet amplitude PDFs at all scales considered are shown in Figure \ref{fig:z6_amps_mock}. 
Note that the scales considered here are slightly different than those at $z=5$ (as listed in table \ref{tab:limit table} and Figure \ref{fig:one_sn_mock_hist}) because the size of the skewers and the nyquist frequency vary at these redshifts. 
At both redshifts we chose 15 logarithmically spaced values for $s_n$ (and $k$). 
At this redshift and signal to noise ratio, we can see distinct bumps for the contribution of noise and wavelet amplitudes for $\SI{290.75}{\kilo\meter\per\second} \geq s_n \geq \SI{55.09}{\kilo\meter\per\second}$. 
Again we have also shown three model wavelet amplitude PDFs: $\log(T_0) = 3.4$ (blue), $\log(T_0) = 4.1625$ (orange), and $\log(T_0) = 4.4$ (green). 
These wavelet amplitude PDFs all agree with each other for the smallest scales $s_n < \SI{23.98}{\kilo\meter\per\second}$ where noise dominates the PDF which is the same for all temperature models. 
The correlation for all wavelet amplitude PDFs at $\log(T_0) = 4.1625$ is shown in Figure \ref{fig:z6_amps_correlation}. 
The $10 \times 10$ diagonal blocks for the largest scales ($s_n > \SI{440.7}{\kilo\meter\per\second}$) and the smallest scales ($\SI{36.34}{\kilo\meter\per\second} > s_n$) look similar to that shown for $z=5$ in Figure \ref{fig:amp_correlation} since the PDFs also have the same shape. 
The $10 \times 10$ diagonal blocks for the mid-range values of $s_n$ are different from the others due to the combined PDF shape as seen in Figure \ref{fig:z6_amps_mock}.
Again the off-diagonal blocks show a similar pattern to the diagonal blocks modified by positive or negative numbers, mimicking the off-diagonal correlations from the power at $z=6$ seen in Figure \ref{fig:z6_power_correlation}. 
The off-diagonal correlation values between the $s_n \leq \SI{23.98}{\kilo\meter\per\second}$ and all other values of $s_n$ show no strong correlations because the uncorrelated noise dominates at these small scales. 

At this redshift we chose not to investigate the combination of the wavelet amplitude PDFs and the power spectrum due to the computational time required and the results at $z=5$ which showed no significant improvement from combining these statistics when the same scales are covered in both.


\bsp	
\label{lastpage}
\end{document}